\documentclass[twocolumn]{aastex631}
\NewPageAfterKeywords
\usepackage{graphicx}% Include figure files
\usepackage{color}
\usepackage{amsmath}  %needed for $\text{}$ and aastex
\usepackage{verbatim}  %for multiline comments with \begin{comment}
\usepackage{hyperref}
\usepackage[noabbrev,capitalise]{cleveref}  %needs to be loaded after hyperref
\usepackage{natbib}
\usepackage[mathscr]{euscript} %script variables: http://www.stat.colostate.edu/~vollmer/pdfs/typesetting-script.pdf
\usepackage{xspace}

\makeatletter
\usepackage{etoolbox}
\patchcmd\H@refstepcounter{\protected@edef}{\protected@xdef}{}{}
\makeatother

\newcommand{\sn}{SN~Ia\xspace}
\newcommand{\sne}{SNe~Ia\xspace}

\newcommand{\ntrans}{ten\xspace}

\newcommand{\h}{\ensuremath{\text{H}_0}\xspace}

\newcommand{\w}{\ensuremath{w_0}\xspace}

\newcommand{\romanFull}{\textit{Nancy Grace Roman Space Telescope}\xspace}
\newcommand{\romanST}{\textit{Roman Space Telescope}\xspace}
\newcommand{\rst}{\textit{Roman}\xspace}

\begin{document}
\title{The Hourglass Simulation:\\A Catalog for the \textit{Roman} High-Latitude Time-Domain Core Community Survey}
\shorttitle{The Hourglass Simulation}

\author[0000-0002-1873-8973]{B. M. Rose}
\affiliation{Department of Physics and Astronomy, Baylor University, One Bear Place \#97316, Waco, TX 76798-7316, USA}
\author[0000-0001-8788-1688]{M. Vincenzi}
\altaffiliation{Einstein Fellow}
\affiliation{Department of Physics, University of Oxford, Denys Wilkinson Building, Keble Road, Oxford OX1 3RH, United Kingdom}
\affiliation{Department of Physics, Duke University, Durham, NC 27708, USA}
\author[0000-0002-0476-4206]{R. Hounsell}
\affiliation{University of Maryland, Baltimore County, Baltimore, MD 21250, USA}
\affiliation{NASA Goddard Space Flight Center, Greenbelt, MD 20771, USA}
\author[0000-0003-1899-9791]{H. Qu}
\affiliation{Department of Physics and Astronomy, University of Pennsylvania, 209 South 33rd Street, Philadelphia, PA 19104, USA}
\author[0000-0003-0183-451X]{L. Aldoroty}
\affiliation{Department of Physics, Duke University, Durham, NC 27708, USA}
\author[0000-0002-4934-5849]{D. Scolnic}
\affiliation{Department of Physics, Duke University, Durham, NC 27708, USA}
\author[0000-0003-3221-0419]{R. Kessler}
\affiliation{Department of Astronomy and Astrophysics, University of Chicago, Chicago, IL 60637, USA}
\affiliation{Kavli Institute for Cosmological Physics, University of Chicago, Chicago, IL 60637, USA}
\author[0000-0002-9946-4635]{P. Macias}
\affiliation{Department of Astronomy and Astrophysics, University of California, Santa Cruz, CA 95064, USA}
\author[0000-0001-5201-8374]{D.~Brout}
\affiliation{Boston University Department of Astronomy, 725 Commonwealth Ave, Boston USA}
% ---
\author[0000-0002-5389-7961]{M. Acevedo}
\affiliation{Department of Physics, Duke University, Durham, NC 27708, USA}
\author[0000-0003-3917-0966]{R.~C.~Chen}
\affiliation{Department of Physics, Duke University, Durham, NC 27708, USA}
\author[0000-0001-6395-6702]{S. Gomez}
\affiliation{Center for Astrophysics, Harvard \& Smithsonian, 60 Garden Street, Cambridge, MA 02138, USA}
\author[0000-0001-8596-4746]{E. Peterson}
\affiliation{Department of Physics, Duke University, Durham, NC 27708, USA}
\author[0000-0001-5402-4647]{D. Rubin}
\affiliation{Department of Physics and Astronomy, University of Hawai`i at M$\bar{a}$noa, Honolulu, Hawai`i 96822, USA}
\affiliation{E.O. Lawrence Berkeley National Laboratory, 1 Cyclotron Rd., Berkeley, CA 94720, USA}
\author[0000-0003-2764-7093]{M.~Sako}
\affiliation{Department of Physics and Astronomy, University of Pennsylvania, 209 South 33rd Street, Philadelphia, PA 19104, USA}
\collaboration{15}{the Roman Supernova Project Infrastructure Team}

\correspondingauthor{B. M. Rose}
\email{Ben\_Rose@baylor.edu}
\shortauthors{The Roman Supernova Cosmology Project Infrastructure Team}

\received{April 4, 2025}
\revised{May 30, 2025}
% \accepted{}
% \reportnum{\href{https://arxiv.org/abs/}{arXiv:3}}
\submitjournal{The Astrophysical Journal}

\begin{abstract}
We present a simulation of the time-domain catalog for the \textit{Nancy Grace Roman Space Telescope}'s High-Latitude Time-Domain Core Community Survey. This simulation, called the Hourglass simulation, uses the most up-to-date spectral energy distribution models and rate measurements for \ntrans extra-galactic time-domain sources. We simulate these models through the design reference \textit{Roman Space Telescope} survey: four filters per tier, a five day cadence, over two years, a wide tier of 19~deg$^2$ and a deep tier of 4.2~deg$^2$, with $\sim$20\% of those areas also covered with prism observations. We find that a science-independent \textit{Roman} time-domain catalog, assuming a S/N at max of $>$5, would have approximately 21,000 Type Ia supernovae, 40,000 core-collapse supernovae, around 70 superluminous supernovae, $\sim$35 tidal disruption events, 3 kilonovae, and possibly \deleted{the first confirmed detection of} pair-instability supernovae. In total, Hourglass has over \replaced{70,000}{64,000} transient objects, \replaced{12}{11}~million photometric observations, and \replaced{560,000}{500,000} spectra.
Additionally, Hourglass is a useful data set to train machine learning classification algorithms. We show that SCONE is able to photometrically classify Type Ia supernovae with high precision ($\sim$95\%) to a $z > 2$. 
Finally, we present the first realistic simulations of non-Type Ia supernovae spectral-time series data from \textit{Roman}'s prism.
\end{abstract}

\keywords{\href{http://astrothesaurus.org/uat/1671}{Surveys (1671)},
\href{http://astrothesaurus.org/uat/205}{Catalogs (205)}, 
\href{http://astrothesaurus.org/uat/2109}{Time domain astronomy (2109)},
\href{http://astrothesaurus.org/uat/1547}{Space telescopes (1547)},
\href{http://astrothesaurus.org/uat/1857}{Astronomical simulations (1857)}
}

\section{Introduction}\label{intro}

The \romanFull is NASA's next flagship mission, with a design focused on answering questions in cosmology, measuring exoplanet demographics, and conducting near-infrared astrophysics surveys \citep{Spergel2015}. It has a $2.4$~meter primary mirror and will be located at the Earth-Sun L2 point. \rst's primary instrument is the Wide Field Instrument (WFI), which has 18 near-infrared detectors (0.48--2.3~$\mu$m and 0.11~arcsec/pix) that create a field-of-view of 0.281~deg$^2$: an expansive 200$\times$ the field-of-view of the \textit{Hubble Space Telescope}'s Wide Field Camera 3 infrared channel.\footnote{\url{https://roman.gsfc.nasa.gov/science/WFI_technical.html}}
NASA has committed to launch the \romanST no later than May 2027.

Up to 75\% of the first five years of \romanST observations will be dedicated to large core surveys. The data set from the \rst High-Latitude Time-Domain Core Community Survey will provide the majority of the extra-galactic transient data from this mission. The observing strategy for this survey is not finalized, however, an initial reference survey has been developed \citep{Rose2021c, Hounsell2023}.\footnote{During the time of this paper was written, the Core Community Survey Definition Committees met and provided a proposed survey definitions. The \rst Time Allocation Committee is currently meeting to finalize the survey.} This reference survey is not a fully optimized design, but rather an example to show that the science goals of \rst are achievable. It is also an excellent baseline for simulations of ancillary science, such as a generic time domain catalog.

The most recent High-Latitude Time-Domain reference survey \citep{Rose2021c,Hounsell2023} will observe one or two discrete locations in the \rst continuous-viewing-zone---within $36^{\circ}$ of either the north or south ecliptic pole---every 5~days over the course of two years. The reference survey includes two filter-depth combinations, referred to as tiers: one tier with $\sim$19~deg$^2$ of $RZYJ$ (0.48--1.454~$\mu$m) at 25.5~mag depth per visit and another 
 $\sim$5~deg$^2$ field with $YJHF$ (0.927--2.00~$\mu$m)  at 26.6~mag depth.
Additionally, there are slitless spectroscopy observations of about 1/10th the total area \citep{Rose2021c,Rubin2022,Wang2023} using an $R\sim100$ prism (0.75--1.80~$\mu$m).

The reference survey is used to demonstrate that \rst can achieve its science goal of understanding dark energy. As such, there is a need to investigate the expected time-domain data set. It is important for the community to understand the capabilities of the \romanST as NASA starts the public process of defining the core community surveys\footnote{\url{https://roman.gsfc.nasa.gov/science/ccs_community_input.html}}. It is valuable to not only release a simulated data set, but also the simulation input files, as this can be used for studying contaminated cosmological samples, and has been requested by the community\footnote{\url{https://pcos.gsfc.nasa.gov/TDAMM/docs/TDAMM_Report.pdf}}.

% Describe what PLAsTiCC did and why.
Simulating a wide range of time-domain sources has previously been done for the Vera C. Rubin Legacy Survey of Space and Time (LSST).
Both the Photometric LSST Astronomical Time Series Classification Challenge  \citep[PLAsTiCC,][]{ThePLAsTiCCteam2018,Kessler2019a} and the Extended LSST Astronomical Time-series Classification Challenge\footnote{\url{https://portal.nersc.gov/cfs/lsst/DESC_TD_PUBLIC/ELASTICC/}} (ELAsTiCC) have produced large simulated data sets for community analysis. These two programs were designed for improving photometric-classification and data-broker pipelines, respectively. 

In this work, we present the Hourglass simulation: a suite of simulated light curves and spectral time series data for \ntrans extragalactic transient types.
Hourglass will allow the community to evaluate the science possible with the current \rst High-Latitude Time-Domain Core Community survey, and assist in the ability to optimize this survey.
In \cref{sec:sim}, we present the assumptions and methods of our simulations. In \cref{sec:data}, we present results of our simulations. Finally, in \cref{sec:conclusions}, we discuss what might be possible in future simulations. 

\section{Simulations}\label{sec:sim}

For this work, we use the simulation components of the SuperNova ANAlysis Software \citep[\texttt{SNANA},][]{Kessler2009} with the \texttt{PIPPIN} \citep{Hinton2020} pipeline manager.
A brief overview of the simulation is as follows:
\begin{enumerate}
    \item Fluxes are generated from an astronomical source spectral-temporal model that starts from a rest-frame spectral model for each epoch. The model is then subject to astrophysical and observational effects such as redshifting. 
    \item The source's spectra are integrated for each filter (or spectral resolution element) to obtain broadband fluxes.
    \item Finally, survey and follow up logic is applied.
\end{enumerate}
The details of our simulations are explained in the following sections, first looking at the sources (\cref{sec:sed}), then summarizing the survey design, selection criteria and other astrophysical parameters (\cref{sec:trigger}), and describing the noise and instrument characteristics (\cref{sec:noise}).

The full set of simulation input files used can be found at \url{https://github.com/Roman-Supernova-PIT/hourglass_snana_sims} and in the data release (\url{https://doi.org/10.5281/zenodo.14262943}).

\subsection{Transients}\label{sec:sed}

In total, we simulate \ntrans transient types as our source models. A list of transient types and their associated parameters can be found in \cref{tab:sim}. We take the spectral energy distributions (SED), absolute magnitudes, and rates from various sources in the literature. For our simulations, we predominately use the publicly released models developed for PLAsTiCC\footnote{\url{https://zenodo.org/record/6672739}} \citep{Kessler2019a,Modelers2022}, though we use several models that have been updated since then. We use the SED, luminosity functions, and luminosity scatter models that comes with each model, unless noted otherwise.

We note that our simulations do not use the library of near-infrared-extended SED models developed for the recently published OpenUnivers2024 \citep{OpenUniverse2025}. This is because the diversity of transient types included in the OpenUnivers2024 transients’ template library is significantly reduced compared to the PLAsTICC transient library. 
Leaving out the near-infrared extensions is only apparent in one or two bands of the lowest redshift objects.

\subsubsection{Type Ia Supernovae}

For the simulation of Type Ia supernovae (SNe~Ia), we use the generative SALT3-NIR model for the SED \citep{Pierel2022}. 
We use the well-tested intrinsic scatter model presented in \cite{Guy2010} along with the population parameters defined in \cite{Kessler2017}. The differences between scatter models \citep[i.e.,][]{Guy2010,Chotard2011,Popovic2022} are negligible ($<$0.1~mag) when not focusing on precise cosmological distances.
SALT3-NIR is defined from 2,800--20,000~\AA.

We use the volumetric rate of:
\begin{equation}\label{eqn:ia-rate}
R(z)= 
    \begin{cases}
        2.4\times10^{-5}\times(1+z)^{1.55}~\text{yr}^{-1}\text{Mpc}^{-3} & \text{if } z < 1.0\\
        7.5\times10^{-5}\times(1+z)^{-0.1}~\text{yr}^{-1}\text{Mpc}^{-3} & \text{if } z \geq 1.0
    \end{cases}  
\end{equation}
as measured in \citet{Strolger2020} thus updating the volumetric rate assumptions used in previous \rst simulations \citep{Hounsell2018,Rose2021c}.
For this simulation, we assume that the 
measured redshift-dependence of the rate from $1<z<2$ can be extrapolated out to $z=3$, as there are no available measurements of SN Ia rates above redshift 2.

For each simulated SN~Ia, the host galaxy is assigned following the SN rates as a function of host galaxy properties published by \cite{Wiseman2021}, similarly to the method presented in \cite{Vincenzi2021}. 
Host galaxies are selected from the 3DHST program \citep{VanDerWel2014}.
Correlations between SN properties and host galaxy properties are modeled following the approach presented by \citep{Popovic2021}.

\subsubsection{SNIa-91bg Supernovae}

SN1991bg-like supernovae (SNIa-91bg) represent the faintest end of the thermonuclear \sn population \citep{Filippenko1992b}. These objects represent about 15\%--20\% of the \sn class \citep{Li2011,Graur2017}. For these simulations, we use a SNIa-91bg model derived in \cite{Gonzalez-Gaitan2014}, which contains 35 templates over a wavelength range of 1,000--12,000~\r{A}. 

We simulate SNIa-91bg supernovae at 15\% of the rate of \sne (see Equation~\ref{eqn:ia-rate}).
The high-redshift rate of SNIa-91bg SN is uncertain since they are preferentially found in more passive and massive galaxies \citep{Gonzalez-Gaitan2011} and a stronger decline in rate at higher redshift is expected. We keep the rate of SNIa-91bg SNe at 15\% of \cref{eqn:ia-rate}, and use the SNIa-91bg-host galaxy correlations assumed in the Dark Energy Survey SN simulations \citep{Vincenzi2021}.

\begin{center}
    \begin{deluxetable*}{llccc}
% \tablecolumns{2}
% \tabletypesize{\small}
\tablewidth{0pt}
\tablecaption{Summary of Transient Input Values\label{tab:sim}}
\tablehead{ 
    \colhead{Transient} &  \colhead{SED Model} & \colhead{SED Wavelengths}  & \colhead{Simulated} & \colhead{Simulated Phase}\vspace{-0.5em}\\
    & \colhead{} & \colhead{(\AA)} & \colhead{Redshifts} & \colhead{}
}   
    \startdata
    SN Ia & \citet{Pierel2022}\tablenotemark{\footnotesize *} & 2,800--20,000 & 
    0.03--2.9 & -40--90\\
    SNIa-91bg & \cite{Gonzalez-Gaitan2014}\tablenotemark{\footnotesize *} & 1,000--12,000 & 
    0.03--2.9 & -40--90\\
    SN Iax & Jha \& Dai 2019 & 1,000--25,000 & 
    0.03--2.9 & -40--90\\
    CCSN & \citet{Vincenzi2019}\tablenotemark{\footnotesize *} & 1,605--11,000 & 
    0.03--2.9 & -40--90\\
    \tableline
    SLSN-I & \texttt{MOSFiT slsn} & 1,000--11,000 & 
    0.03--2.9 & -50--150\\
    TDE & \texttt{MOSFiT tde} & 1,000--11,000 & 
    0.03--2.0 & -40--90\\
    ILOT & \texttt{MOSFiT csm} & 1,100--11,000 & 
    0.03--1.0 & -20--25\\
    KN & \cite{Bulla2019} & 100--99,900  & 
    0.03--1.5 & -2--20\\
    PISN & \texttt{MOSFiT default} & 1,000--11,000 & 
    0.03--2.9 & -50--150\\
    \hline
    AGN & ELAsTiCc, in prep. & 100--20,000 & 0.03--2.9 & -40--90 \\
    % Non-variable &\nodata & \nodata & \nodata & -30--30 \\
    \enddata
\tablenotetext{*}{These models have been updated since PLAsTiCC to extend their wavelengths into the rest frame near-infrared.}
\end{deluxetable*}
\end{center}

\subsubsection{Type Iax Supernovae}

Type Iax supernovae \citep[or 02cx-like,][hereafter SNe~Iax]{Li2003} are thought to be failed thermonuclear runaways \citep{Foley2013,Jha2017}. 
For this simulation, we use templates created by S. Jha and M. Dai\footnote{\url{https://github.com/RutgersSN/SNIax-PLAsTiCC}} that are based on observations of SN~2005hk and are adapted to match the luminosity function and light-curve parameters for SN~Iax.
The 1000 SN~Iax templates have a rest-frame wavelength range of 1,000--25,000~\r{A} and include A$_{\mathrm{V}}$ variations \citep{Vincenzi2021}.

The volumetric rate for SN~Iax is set by local measurements relative to the \sne rate \citep[30\%,][]{Foley2013,Miller2017}. With this information, we simulate SN Iax at 30\% of the SNIa volumetric rate (see Equation~\ref{eqn:ia-rate}), and use the transient-host galaxy correlations assumed in the Dark Energy Survey SN simulations \citep{Vincenzi2024}.

\subsubsection{Core-collapse Supernovae}

For core-collapse supernovae (Type Ib/c and Type II, hereafter referred to as CCSN) we use 65 spectral models from \cite{Vincenzi2019}, including the CCSN-host galaxy property correlations.
We use the rates derived from the GOODS, CANDELS, and CLASH surveys \citep{Strolger2015}. This rate is 
\begin{equation}\label{eqn:Strolger2015}
    R(z) = 0.015\frac{(1+z)^{5}}{((1+z)/1.5)^{6.1} + 1} ~\text{yr}^{-1}\text{Mpc}^{-3} ~~~.
\end{equation}
\added{This rate was only measured out to $z \sim 2.5$, therefore we make the assumption that it is valid to $z = 3$. Additionally, each parameter in this equation has an uncertainty of $\sim$10\%.}

\subsubsection{Superluminous Supernovae}

Superluminous supernovae type I (SLSN-I) are some of the most luminous transients, with a peak absolute magnitude of $\lesssim$$-21$~mag. They are fairly rare and only recently discovered \citep{Quimby2011,Chomiuk2011}. For this simulation, we use the Modular Open-Source Fitter for Transients \citep[\texttt{MOSFiT},][]{Guillochon2018} generated spectral-temporal SEDs using the 
\texttt{MOSFiT slsn} model that assumes a magnetar engine and a blackbody SED \citep{Kasen2010a,Nicholl2017}. 
This model is defined from 1,000--11,000~\r{A} and resulted in 960 templates.

We use the SLSN rate from \cite{Prajs2017}:
\begin{equation}
    R(z) = 2.0\times10^{-8}\times R_{\text{MD14}}(z) ~~~.
\end{equation}
where $R_{\text{MD14}}(z)$ is the redshift dependent component of the cosmic star formation rate defined in \cite{Madau2014} and repeated here:
\begin{equation}
    \label{eqn:MD14}
    R_{\text{MD14}}(z) = \frac{(1+z)^{2.7}}{((1+z)/2.9)^{5.6} + 1} ~\text{yr}^{-1}\text{Mpc}^{-3} ~~~.
\end{equation}
This rate is a slight variation to what was reported in \cite{Madau2014}. We do not use the multiplicative scaling of $0.015$ that is seen in Equation 15 of \cite{Madau2014} since we want to normalize each astrophysical object to its own $z=0$ rate.
This SLSNe rate is within the current uncertainty presented in the literature \citep{Quimby2013,McCrum2015,Prajs2017}, and reflects the fact that the SLSNe rate appears to follow the cosmic star formation rate \citep{Prajs2017}. For the SLSN, we use the transient-host galaxy correlations derived for the OpenUniverse2024 simulations \citep{OpenUniverse2025}. These are built off of the work of the Simulated Catalog of Optical Transients and Correlated Hosts \citep[SCOTCH,][]{Lokken2023}, but extended into the \rst wavelength range.

\subsubsection{Tidal Disruption Events}

Tidal disruption events (TDEs) occur when stellar objects are tidally ripped apart as they fall into a black hole. Typically, this takes place near supermassive black holes due to the required orbital-to-stellar radius ratio \citep{Rees1988,Mockler2019}. We use SEDs generated from the \texttt{MOSFiT tde} model and transient-host galaxy property correlations derived for the OpenUniverse2024 and SCOTCH simulations \citep[][respectively]{OpenUniverse2025,Lokken2023}.

We use the TDE rate from \cite{Kochanek2016} and \cite{vanVelzen2018}: 
\begin{equation}
    R(z) = 10^{-6}\times10^{-5z/6}~\text{yr}^{-1}\text{Mpc}^{-3} ~~~.
\end{equation}
\added{We assume this rate holds throughout cosmic time, however, \cite{Kochanek2016} argues that the rate of TDE's sharply drops off at $z=1$. This claim is something \romanST can test.}

\subsubsection{Intermediate Luminosity Optical Transients}

Intermediate luminosity optical transients (ILOTs) have peak luminosities between classical novae and supernovae and display signs of interaction with a dense circumstellar material, similar to SNe IIn.
The 385 templates we use were generated using the \texttt{MOSFiT csm} model following parameter ranges described in \cite{Villar2017} and \cite{Chatzopoulos2012} with a rest-frame wavelength range of 1,000--11,000~\r{A}. %http://adsabs.harvard.edu/abs/2017ApJ...849...70V
We assume a volumetric rate that is 6\% that of the CCSN rate defined in \cref{eqn:Strolger2015}, and use the CCSN-host galaxy correlations from the Dark Energy Survey. \added{This rate is from \citet{Li2011a} following the assumption that ILOTs have a similar progenitor as Type IIn supernovae.}

\subsubsection{Kilonovae}

Kilonovae (KN) are the merger of two compact objects, typically neutron stars \citep{Kasen2017}. 
We use the SED model defined in \cite{Bulla2019}. 
Our model has 550 templates with a rest-frame wavelength range of 100--99,900~\r{A}.
We use the volumetric rate of
\begin{equation}\label{eqn:kn-rate}
    R(z) = 3.2\times10^{-7}~\text{yr}^{-1}\text{Mpc}^{-3}
\end{equation}
from a recent estimate of the binary neutron stars \citep{Abbott2021}. However, this rate is very uncertain, and the total number of KN observed by \romanST could be over or underestimated by a factor of $\sim$3 \citep{Abbott2021}.
Additionally, to reduce Poisson noise in the detection rate, we scale up these simulations by using a rate five times that of \cref{eqn:kn-rate} to ensure the total detected objects is $>$10.

\subsubsection{Pair Instability Supernovae}

Pair instability supernovae (PISN) are proposed to arise when low-metallicity Population III stars, with M$_{\star} \sim 140\text{--}260~\mathrm{M}_{\odot}$, reach sufficiently high core temperatures that $\gamma$-rays produce electron-positron pairs \citep{Barkat1967,Kasen2011}. 
PISN are extremely high redshift transients, and therefore only a few possible PISN have been reported \replaced{\citep{Gal-Yam2009,Cooke2012,Kozyreva2018}}{\citep{Gal-Yam2009,Cooke2012,Kozyreva2018,Gomez2019,Schulze2024,Aamer2024}}.
We use the \texttt{MOSFiT default} model with PISN parameters as described in \cite{Villar2017}.
We use 1000 SED time series templates that have a rest-frame wavelength range of 
1000--11000~\r{A}.

A volumetric rate for PISN presented stated in \cite{Briel2022}. We fit a polynomial to the data presented and obtain a volumetric rate of 
\begin{multline}
    R_{B22}(z) =
     [-4.08 + 30.56z - 19.91z^2 \\ +16.82z^3 - 3.00z^4]~\text{yr}^{-1}\text{Gpc}^{-3} ~~.
\end{multline}
At $z=1$ it is nearly twice the PISN rate \citep{Pan2012} used in previous simulation efforts.
Since PISN are from extremely massive stars, we use the CCSN-host galaxy correlations derived from the Dark Energy Survey.

\subsubsection{Active Galactic Nuclei}

We simulate 500 active galactic nuclei (AGN) that have a constant volumetric rate \added{($1.0^{-3}~\mathrm{Mpc}^{-3}$)} for the redshift range $0.03 < z < 3$. We use the analytical model that has been generated for the ELAsTiCc simulations. It is a damped random walk model that includes the possibility of changing look behavior. The simulation varies the black hole mass, Eddington ratio \citep{Weigel2017,Sartori2019}, and $i$-band absolute magnitude \citep{Shen2013} for each object. This model will be described in further detail in the ELAsTiCc paper in prep.

\subsubsection{Non-Variable Sources}

To allow for the widest range of tests, we include ``fixed magnitude sources.'' These are objects that have the same magnitude at all epochs, but a different random magnitude is chosen for each event. We simulate 1000 events that range in apparent magnitude from 18 to 30~mag.

\subsection{Survey Design and Detection}\label{sec:trigger}

We follow the ``25\% spectroscopic time'' survey strategy laid out in \cite{Rose2021c}, where the wide tier is observed with the F062, F087, F106, and F129 (colloquially $R$, $Z$, $Y$, $J$) and covers 19.04~deg$^2$. The deep tier is observed in F106, F129, F158, and F184 ($Y$, $J$, $H$, $F$) and covers 4.20~deg$^2$. These exposure times and areas were determined when the average slew and settle time would have been $\sim$70~s. This has since been reduced, and the expectation is that the actual survey should have more ``open-shutter'' time. For these simulations, we keep the 70~s slew and settle time assumption.
With many of the SED models only extending redward to 11,000~\r{A}, the simulations for these events only have $R$, $Z$, and $Y$ until $z>0.175$ where there is full coverage of the $J$-band.
We locate the two tiers such that they do not overlap and are located near Euclid's Deep Field South, roughly (RA, dec) = (04:04:58, $-$48:25:23). 
We also simulate the slitless prism with 12 wide-tier and 4 deep-tier pointings (3.36 and 1.12~deg$^2$ respectively).
All exposure times and numbers of pointings per tier are taken from \cite{Rose2021c} and summarized in \cref{tab:exposures}. 
We have designed these simulations so that they can be updated once an observational strategy has been selected by the \rst scientific community.
\begin{deluxetable}{lccr}
% \tablecolumns{2}
% \tabletypesize{\small}
\tablewidth{0pt}
\tablecaption{Summary of Observing Strategy from \cite{Rose2021c}\label{tab:exposures}}
\tablehead{ 
    \colhead{Filters} & \colhead{Exp. Time} & \colhead{No. of} & \colhead{Area}
    \vspace{-0.5em}\\
    \colhead{} & \colhead{(s)} & \colhead{Pointings} & \colhead{(deg$^2$)}
} 
\startdata
    \sidehead{\hspace{-1.5em}\textit{Wide Tier}}
    RZYJ & 160;100;100;100 & 68 & 19.04 \\ 
    prism & \hfill 900 ~~~~ & 12 & 3.36 \\ 
    \sidehead{\hspace{-1.5em}\textit{Deep Tier}}
    YJHF & 300;300;300;900  & 15 &  4.20 \\ 
    prism & \hfill 3600~~~~ &  4 & 1.12 \\ \enddata
\end{deluxetable}

\added{The Roman Time Allocation Committee (ROTAC) recently released recommendations for the Core Community Surveys \citep{ROTAC2025}. This is similar to the design reference mission but increases the area of the deep tier (from 15 to 23 pointing) and adds additional filters at the cost of using an interweaving cadence (rotating sets of filters so an observation is still taken every 5~days but the filter to same filter cadence is 10~days). As a result, the expected number of objects increases by approximately 25--30\%. Additionally, they recommended a pilot and extended surveys outside the core, two-year, survey. We plan on updating these simulations when the a stable version of the ROTAC recommendations are implemented, likely Fall~2025.}

Additionally, these simulations do not account for dithering, roll angles or field edge effects. Some ramifications of these effects have been discussed previously in \cite{Rose2021c} and \cite{Hounsell2023}. For current approximations of these effects, one can discard one of every eight photometric observations, representing \romanST's $\sim 7/8$ths focal plane fill factor.

Regarding detection, we impose a threshold of two observations with S/N$>$5.\footnote{This cut is on single epoch images. Presumably one can stack epochs and increase the total detections for slow transients, such as PISN and SLSN.}
This is a loose quality cut that only removes the lowest signal to noise simulated events. It is also a fairly standard cut applied to real data in order to remove subtraction artifacts (which are not included in our simulations). This approach of only applying a loose cut in our simulations will allow users to implement their own quality cuts.
We do not apply any selection effects from spectroscopic follow-up.

We use host-galaxy characteristics from the 3DHST program \citep{VanDerWel2014}. Our sample contains 110,798 entries with galaxy properties derived from real data\footnote{\url{https://github.com/srodney/romansims/blob/main/romansims/catalogs.py}}.
This set of host galaxies were used in the pixel simulations found in \cite{Wang2023}.
However, this choice limits our simulations to a redshift maximum of 3. 
This upper limit doesn’t affect the majority of transients included in the simulations, except for SLSNe and PISN, which in principle could be detected far beyond this redshift with \rst. In the future, we plan to update the host galaxy catalogs used for our simulations using N-body simulations such as the Deep Realistic Extragalactic Model (DREaM) Galaxy Catalog \citep{Drakos2022}.

\added{For host galaxy dust, we use an exponential plus gaussian ``galactic line-of-sight" model for $A_V$ presented in \citet[Equation 2]{Wood-Vasey2007}. The Gaussian core has a standard deviation of $\sigma = 0.6$~mag and the exponential tail has a rate parameter $\tau = 1.7$~mag. These values are taken from \cite{Vincenzi2021}.
For $R_V$, we simulate a range of values, 3.0--3.2, peaking at 3.1.
% For two transients we use something different.
For \sn, the host-galaxy dust is already a part of the SALT2 and population models, therefore we don't include more.
% Since TDEs are nuclear transients, we only use the Gaussian component.
}

We use the Milky Way color law from \cite{Fitzpatrick1999} assuming an $R_V = 3.1$ and a dust map from \cite{Schlafly2011}.
In this simulation, we assume a flat-$\Lambda$CDM cosmology with $\Omega_{\text{M}} = 0.3$ and $h=0.7$.

\subsection{\rst WFI Characteristics}\label{sec:noise}

\begin{figure*}
    \centering
    \includegraphics[width=0.45\textwidth]{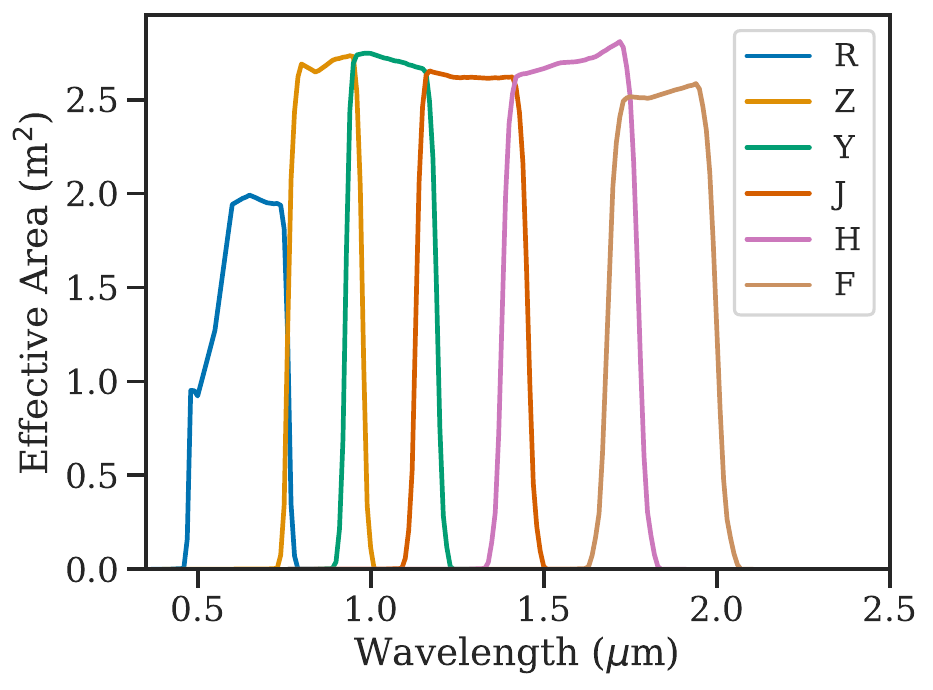}
    \hspace{0.2in}
    \includegraphics[width=0.45\textwidth]{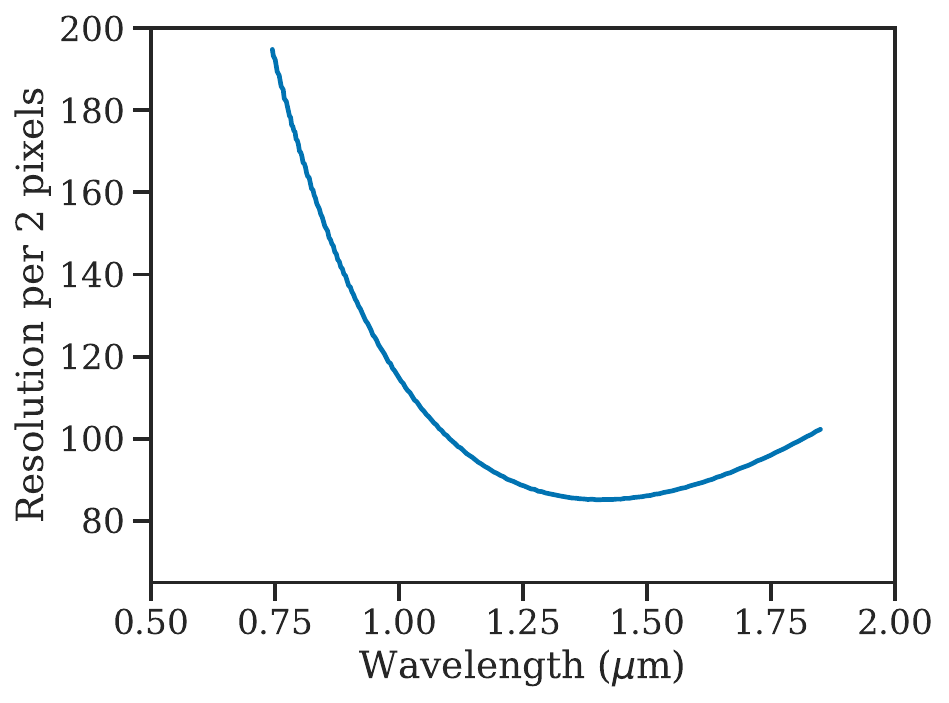}
    
    \caption{(\textit{left}) The filter transmission and detector response functions used for these simulations. Curves are taken from the Roman project technical resources.
    (\textit{right}) The prism's resolution (per 2 pixels) as a function of wavelength. The prism is transmissive from 7,500--18,000~$\mathring{\mathrm A}$. The minimum two-pixel dispersion is $\sim$80 around 14,000~$\mathring{\mathrm A}$. Except for the blue edge, most of the wavelengths have an $R\leq100$.
    }
    \label{fig:filters}
\end{figure*}

The latest hardware characteristics for the \romanST's WFI can be found at \url{https://roman.gsfc.nasa.gov/science/WFI_technical.html}. This URL contains the current estimates for the filter transmission, detector efficiency, prism two-pixel dispersion, and filter throughput. These parameters are incorporated into the simulation's noise model by the creation of appropriate definitions of filter and prism zero-points and point-spread functions. \Cref{fig:filters} (\textit{left}) shows the filter transmission functions used and \cref{fig:filters} (\textit{right}) shows the prism's wavelength dependent two-pixel dispersion. 

\begin{deluxetable*}{lcccccc}
% \tablecolumns{2}
% \tabletypesize{\small}
\tablewidth{0pt}
\tablecaption{Roman WFI Characteristics Used in Simulation\label{tab:noise}}
\tablehead{ 
    \colhead{Filters} & \colhead{PSF$_{\mathrm{NEA}}$} & \colhead{$\sigma_{\mathrm{read}}$} & 
    \colhead{$\sigma^2_{\mathrm{thermal}}$} & 
    \colhead{$\sigma^2_{\mathrm{zodi}}$} & 
    \colhead{$\sigma_{\mathrm{sky}}$} & \colhead{ZP$_{\mathrm{AVG}}$}
    \vspace{-0.5em}\\
    \colhead{} & \colhead{(pix)} & \colhead{(e$^-$/pix)} &
    \colhead{(e$^-$/s/pix)} & 
    \colhead{(e$^-$/s/pix)} & 
    \colhead{(ADU/pix)} & \colhead{(mag)}
} 
\startdata
    \sidehead{\hspace{-1.5em}\textit{Wide Tier}}
     R & 5.575 &  9.011  & 0.003 & 0.315 & 7.133 & 32.129  \\ 
     Z & 6.695 & 10.624  & 0.003 & 0.316 & 5.648 & 31.303  \\
     Y & 7.895 & 10.624  & 0.003 & 0.357 & 6.000 & 31.356  \\
     J & 9.210  & 10.624 & 0.003 & 0.359 & 6.017 & 31.354  \\ 
    \sidehead{\hspace{-1.5em}\textit{Deep Tier}}
     Y & 7.895  & 7.450 & 0.003 & 0.357 & 10.392 & 32.549  \\
     J & 9.210  & 7.450 & 0.003 & 0.359 & 10.421 & 32.547  \\
     H & 11.140 & 7.450 & 0.048 & 0.339 & 10.775 & 32.570  \\
     F & 16.335 & 5.942 & 0.155 & 0.194 & 17.723 & 33.299  \\
     \enddata
\tablecomments{
% READ NOISE: Read noise in photo-electrons, per pixel.
% SKYSIG: Standard deviation of sky, in ADU per pixel.
$N_{\mathrm{read}}$, $N_{\mathrm{sky}}$, and ZP$_{\mathrm{AVG}}$ assume the exposure times listed in \cref{tab:exposures}.
% ZPTAVG: Zero point relating the source magnitude to the flux measured in e$^-$.
}
\end{deluxetable*}
%old sky values
%3.985
%3.160
%3.570
%3.590
%-
%6.184
%6.218
%5.930
%4.301

For the SNANA simulations, the noise values (thermal, sky, and read noises) are calculated to include exposure times. Thus, we have different noise and zero-points per filter and per tier. 
Our values are presented in \cref{tab:noise} and are updated from those assumed in the simulations presented by \cite{Hounsell2018}.
Since these simulations are based on the current reference survey design \citep{Rose2021c}, we use the same exposure times (see \cref{tab:exposures}) and average slew and settle time (70~s).  
However, other instrument properties have been updated since then. The are available in \cite{Mosby2020} or on the WFI technical specifications website, \url{https://roman.gsfc.nasa.gov/science/WFI_technical.html}. 
WFI underwent thermal-vacuum testing in the first half of 2024. We have set up our simulations such that we can update them with post-thermal-vacuum values when they become available.

\subsubsection{Point Spread Function}
We take the point spread function (PSF) from the \rst WFI technical website mentioned above. The PSF has a large variation across the field of view, therefore, we use the median of the best and worst noise equivalent areas (NEA). The values are presented in \cref{tab:noise}. Though we only have one area per filter, we account for the significant NEA change across the field-of-view through the variation of the zero point. This does not fully match the effects of a changing PSF (e.g., it has no dependence on source color), but it does approximate the changes in signal to noise as a point source gets spread over different areas.

\subsubsection{Zero-point Values} 

The simulations use an exposure time-dependent zero-point defined as:
\begin{equation}
    \mathrm{ZP_{\mathrm{AVG}}} = \mathrm{ZP} + 2.5\log_{10}(t)
\end{equation}
with ZP being the magnitude that results in one count per second and $t$ being the total exposure time. ZP was calculated by using the filter effective area curves provided available at the WFI technical specifications website and integrating an flat-AB spectrum. There is a significant scatter in zero-point and PSF across the field of view. To approximate these effects, we include a zero-point scatter with a standard deviation of 0.15~mag. This is a stochastic effect that depends on where a source lands on the field of view.

% ramp read noise -> CCD Noise
\subsubsection{Read Noise} 
The simulation uses the read noise as a function of exposure time. Since near-infrared detectors use non-destructive reads, and \rst will not download every resultant that is read \citep{Casertano2022,Gomez2023}, the total read noise as a function of exposure time can be defined as
\begin{equation}
    \sigma_{\mathrm{read}} = \sqrt{\sigma^2_{\mathrm{floor}} + \sigma^2 \frac{(n-1)}{n(n+1)}}
\end{equation}
where $n = t_{\mathrm{exp}}/3.04$~s or number of reads for a given exposure ($t_{\mathrm{exp}}$). 
We use $\sigma^2_{\mathrm{floor}} = 25$ and $\sigma^2 = 12*16^2 = 3072$ as defined by measurements reported by the project in 2021.\footnote{\url{https://roman.gsfc.nasa.gov/science/RRI/Roman_WFI_Reference_Information_20210125.pdf}}
These values are given from the project on the WFI technical specifications website.
This is the same equation for read noise as used in \cite{Hounsell2018}, but the final numbers are reduced slightly.
The final read noise per filter for our given exposure times can be seen in \cref{tab:noise}.

\subsubsection{Sky Noise}
The simulation combines all sky noise together---zodiacal, and thermal---for a given exposure length. 
The sky-noise is reported as a standard deviation in ADU per pixel.
This can be calculated with the following equation:
\begin{equation}
    \sigma_{\mathrm{sky}} = \sqrt{t_{\mathrm{exp}}(\sigma^2_{\mathrm{zodi}} + \sigma^2_{\mathrm{thermal}})}
\end{equation}
where $t$ is exposure time, and $\sigma^2_{zodi}$, and $\sigma^2_{thermal}$ are given by the project. The dark current is sufficiently small to be ignored this simulation.
These values have reduced since \cite{Hounsell2018}.

\section{Light Curves and Spectral Time Series}\label{sec:data}
\vspace{-2em}
\begin{center}
\begin{deluxetable}{lccc}
% \tablecolumns{2}
% \tabletypesize{\small}
\tablewidth{0pt}
\tablecaption{Hourglass Catalog Summary\label{tab:results}}
\tablehead{ 
    \colhead{Transient} & \colhead{Total Detected} & \colhead{Median S/N}  & \colhead{Median} \vspace{-0.5em}\\
    \colhead{} & \colhead{Detected} & \colhead{at Maximum} & \colhead{Redshift}
}
\startdata
    SN~Ia & 21,700 & 13.5 & 1.32 \\
    SNIa-91bg & \added{1,300} & 10.6 & 0.84 \\
    SN~Iax & \added{1,300} & 8.5 & 0.95 \\
    CCSN & \added{39,000} & 8.8 & 0.90 \\
    \hline
    SLSN-I & 70\tablenotemark{\footnotesize a} & 32.4 & 1.82 \\
    TDE & 39 & 13.5 & 0.65 \\
    ILOT & 35 & 6.7 & 0.49 \\
    KN\tablenotemark{\footnotesize b} & 14 & 7.9 & 0.35 \\
    PISN & 15\tablenotemark{\footnotesize a} & 8.2 & 2.23 \\
    \hline
    AGN & 139 & 13.6 & 1.78 \\
\enddata
\tablecomments{
Detected values are rounded according to the expected counting statistics uncertainty. \added{The ROTAC recommend survey should increase these numbers by 25--30\% for the core survey. Including the pilot and extended surveys, we expect over 100,000 transients.}
\tablenotetext{a}{One could used multi epoch stacked images and increase the total detections for slow transients.}
}
\tablenotetext{b}{This simulated KN rate is five times that of \cite{Abbott2021} to overcome Poisson noise, which has stated uncertainty up to a factor of four.}

\end{deluxetable}

\end{center}

\begin{figure*}
    \centering
    \includegraphics[width=0.47\textwidth]{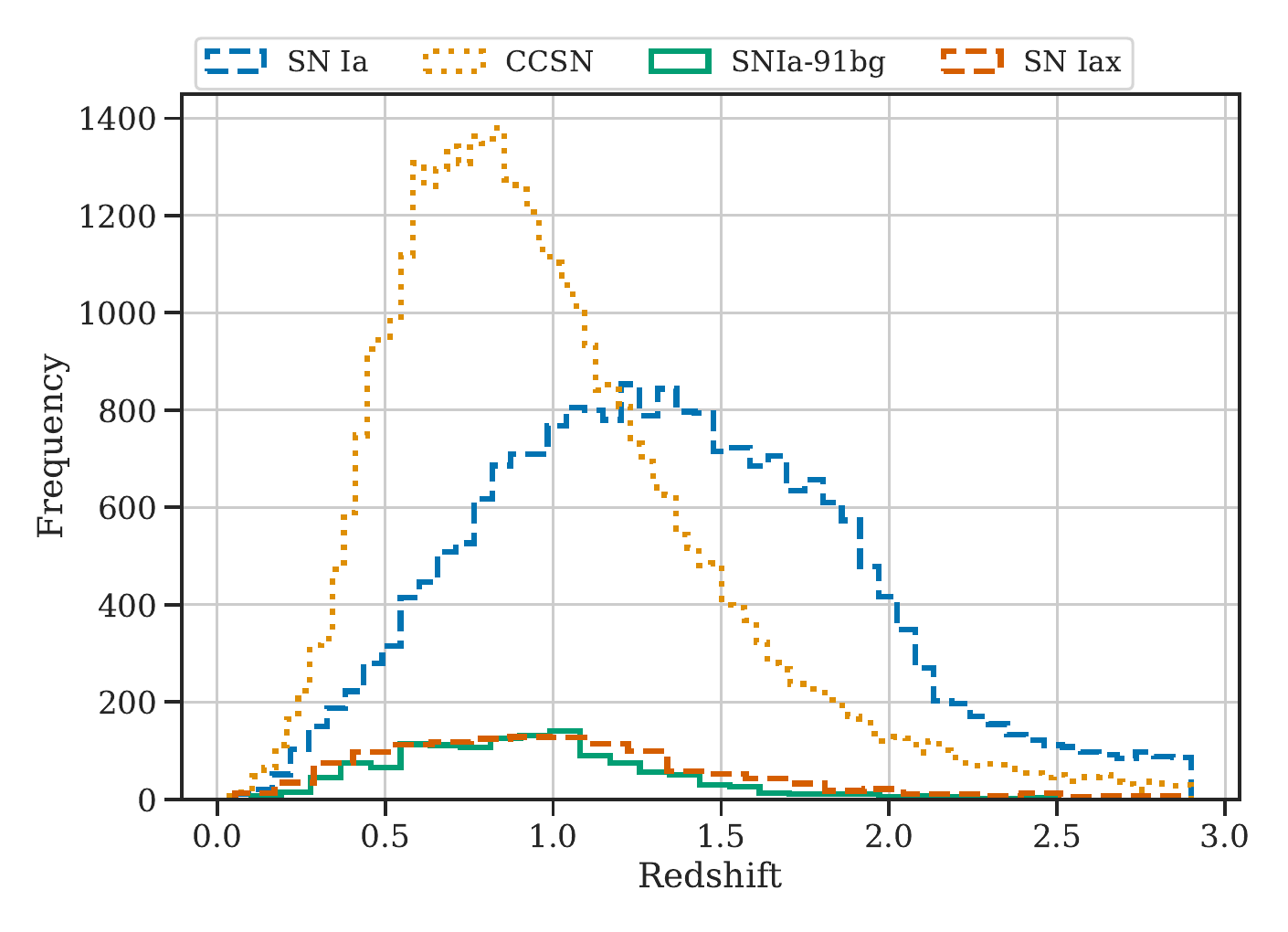}
    \hspace{0.2in}
    \includegraphics[width=0.47\textwidth]{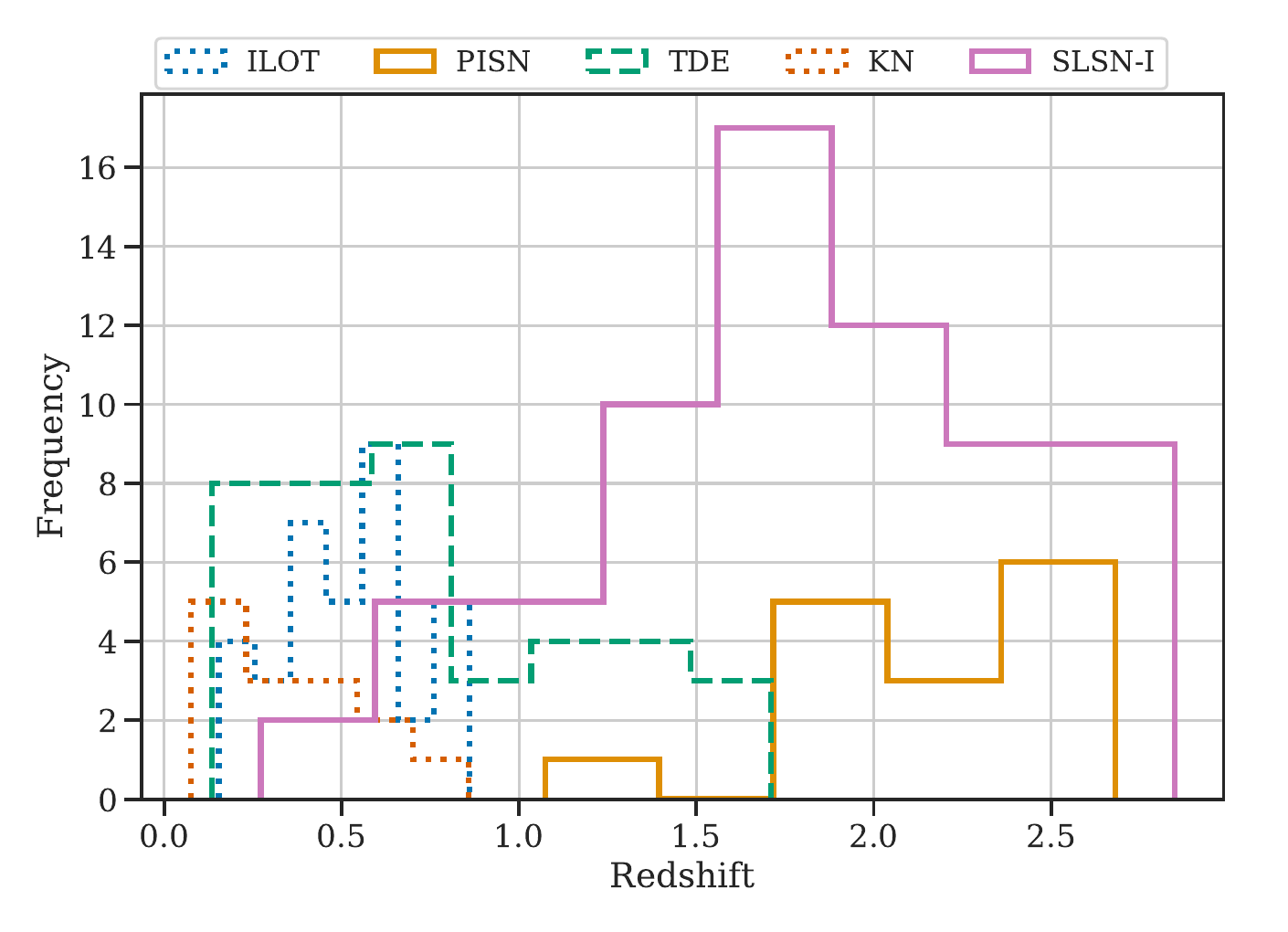}
    \caption{Redshift distribution of recovered transients. The 
    (\textit{left}) panel shows the redshift distributions for high-volume transients such as CCSNe and SNe~Ia, as well as for the fainter SN-Ia-like objects (SNe~Iax and SNIa-91bg). The
    (\textit{right}) panel shows the redshift distribution of rarer transient events, including SLSNe, TDEs, ILOTs, KNe, and PISNe. These distributions are more uncertain due to the limited data used in rate models, particularly at redshifts that will be probed by \rst.
    We do not present the redshift distributions for the AGN in this figure since we generated a fixed number of events rather than use a redshift dependent volumetric rate. The light curves and spectra are still in the data release.
    }
    \label{fig:n}
\end{figure*}

We provide a summary of the Hourglass simulation in \cref{tab:results}.  We give the median redshift for each type and find it varies as expected depending on the intrinsic magnitude.  We show redshift histograms for the different types of transients in \cref{fig:n}.
Along with this paper, we release the photometry, spectroscopy, and object characterization data as parquet files at \doi{10.5281/zenodo.14262943}. Details on this data release are described in Appendix~\ref{data_release}.

We find that SNe Ia, PISN and SLSN can be discovered up to the redshift limit of our simulations at $z=3$. In particular, significantly higher redshift PISN and SLSN will be visible, but this simulation artificially cuts off at $z=3$. \added{\citet{Moriya2022} estimated around 100 PISN at $z>5$ and about a dozen SLSN at the same redshift range. Due to our redshift range, our simulations are not directly comparable but these values are not inconsistent since Hourglass is only seeing the ``low-$z$'' tail of the PISN and SLSN distributions.}

For each class of transient we see the expected dropoff in number due to selection effects (\cref{fig:n}). This effect is most obvious in SNe~Ia and CCSNe. Additionally, we agree with \cite{Inayoshi2024} that \rst will observe 10--60 TDEs, though our number suggests the higher side of that range. 

\begin{figure*}
    \centering
    \gridline{
              \fig{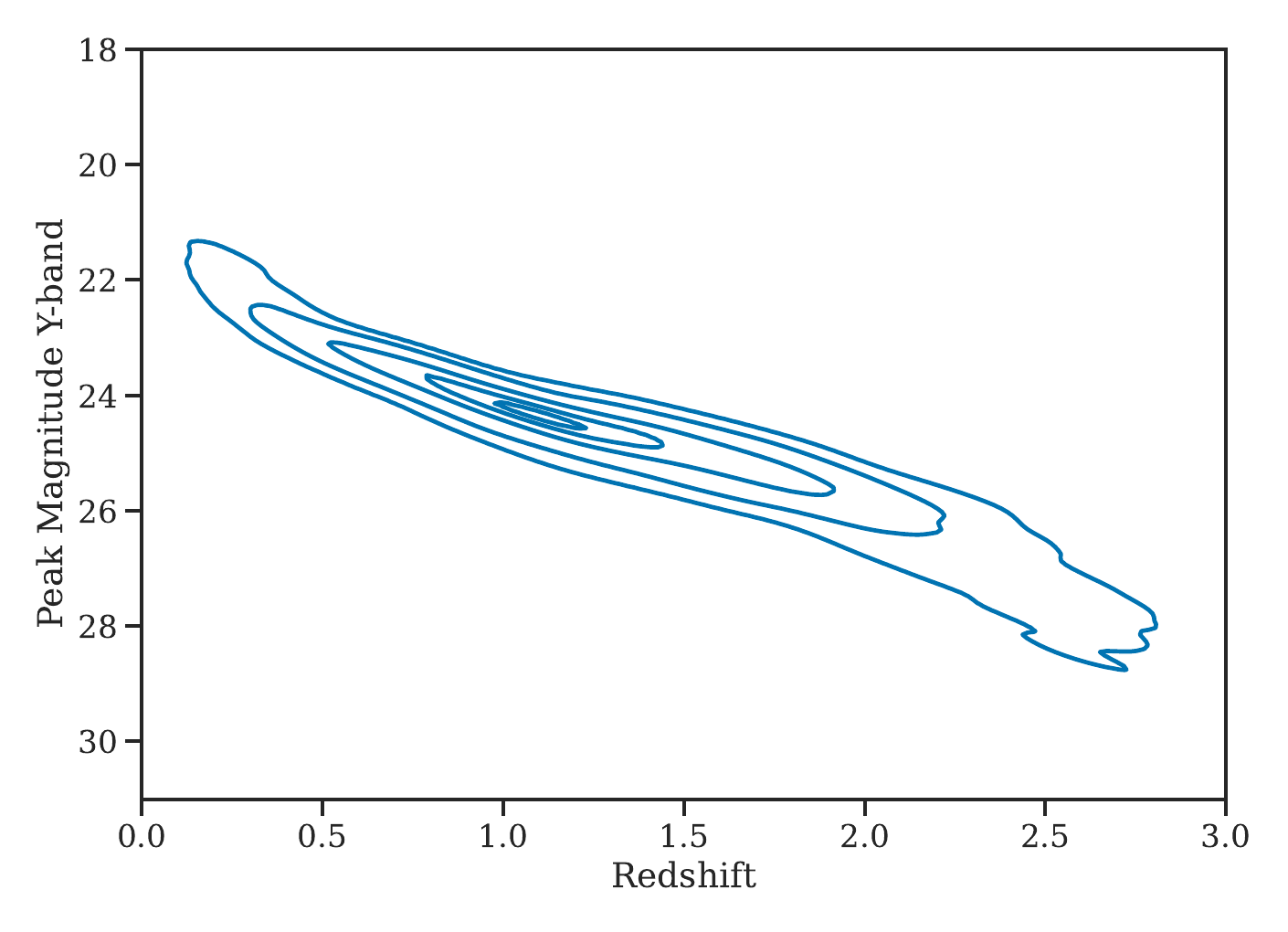}{0.33\textwidth}{SN~Ia}
              \fig{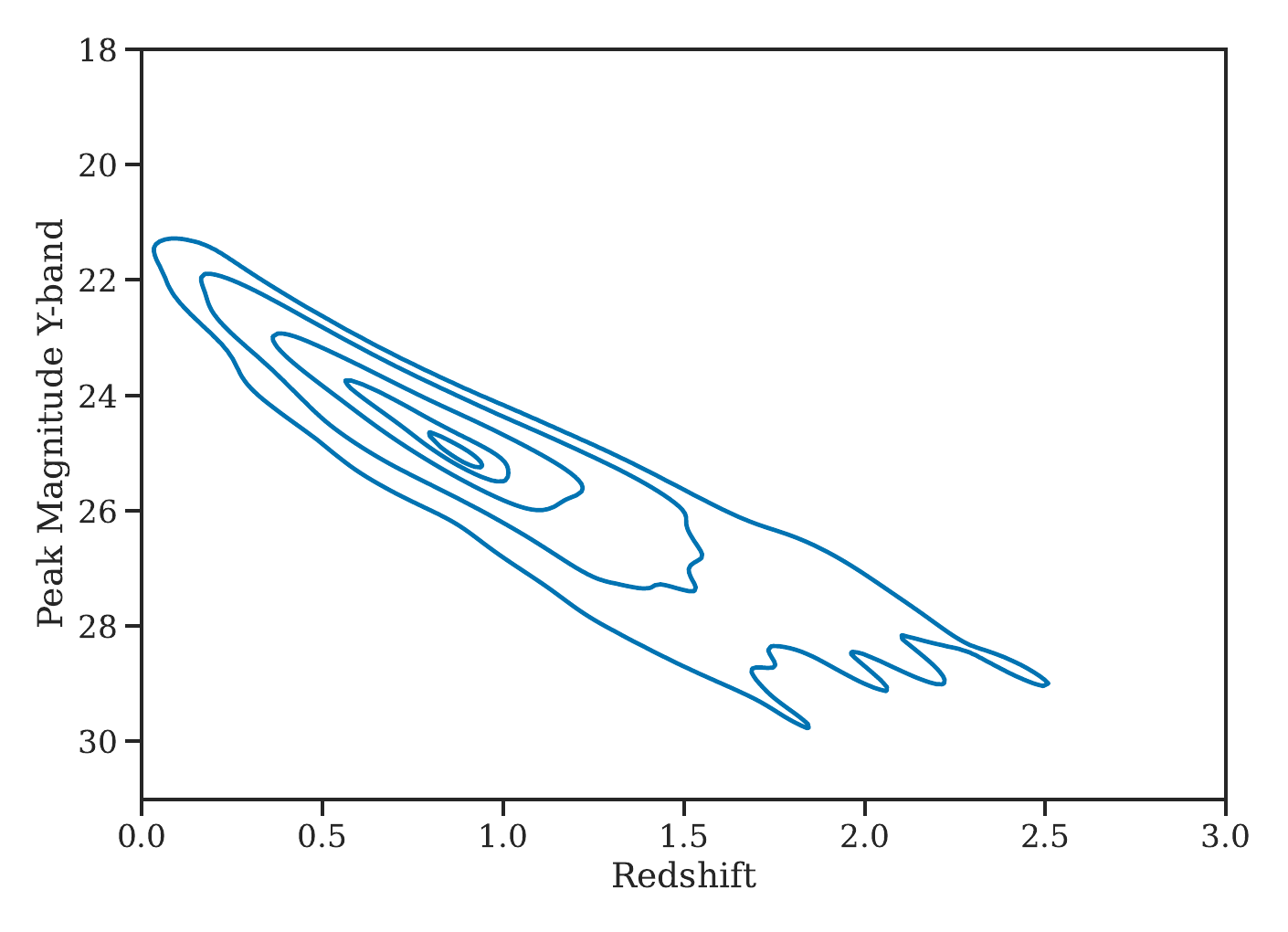}{0.33\textwidth}{SNIa-91bg}
              \fig{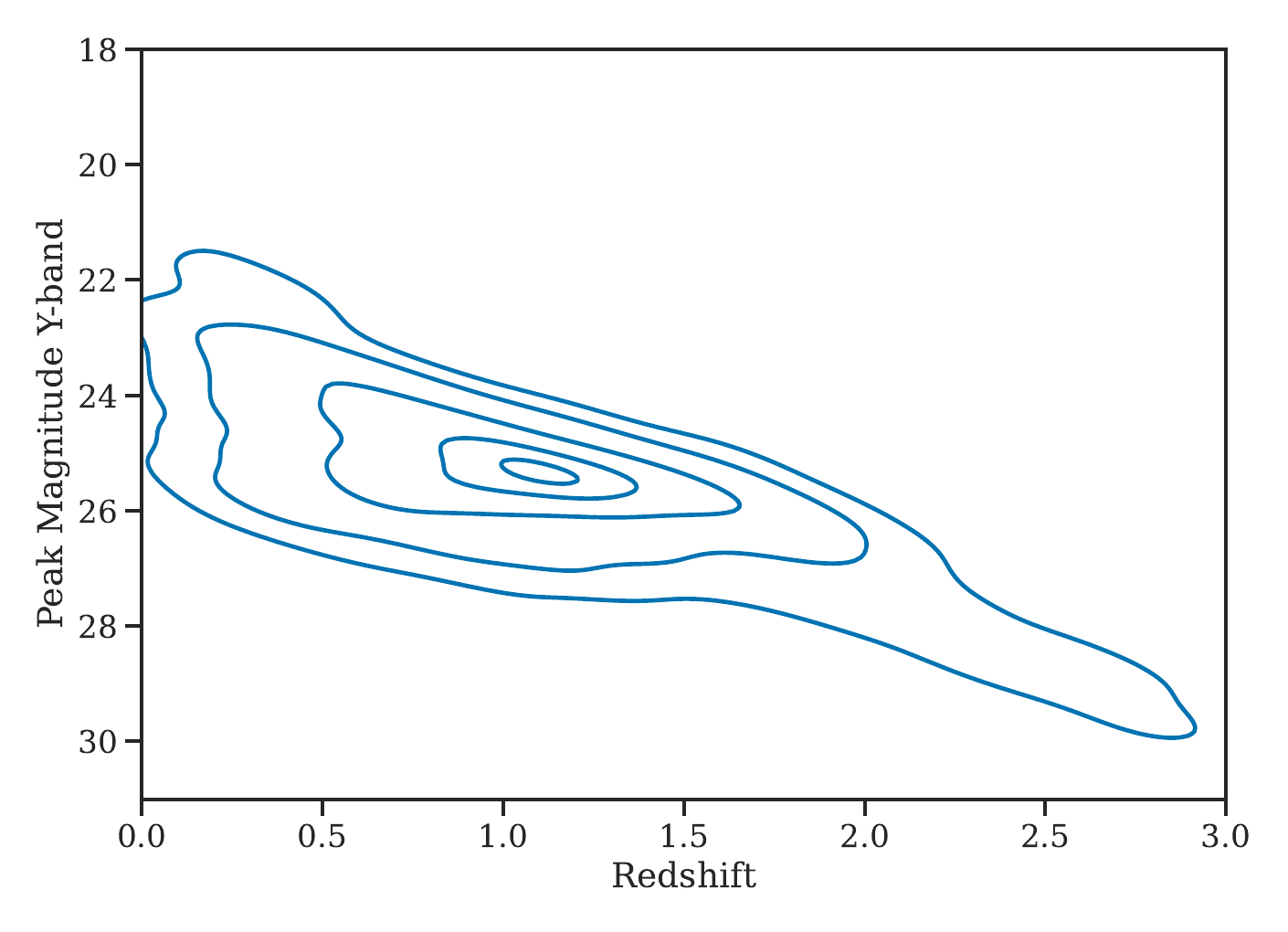}{0.33\textwidth}{SN~Iax}
              }  
    \gridline{
               \fig{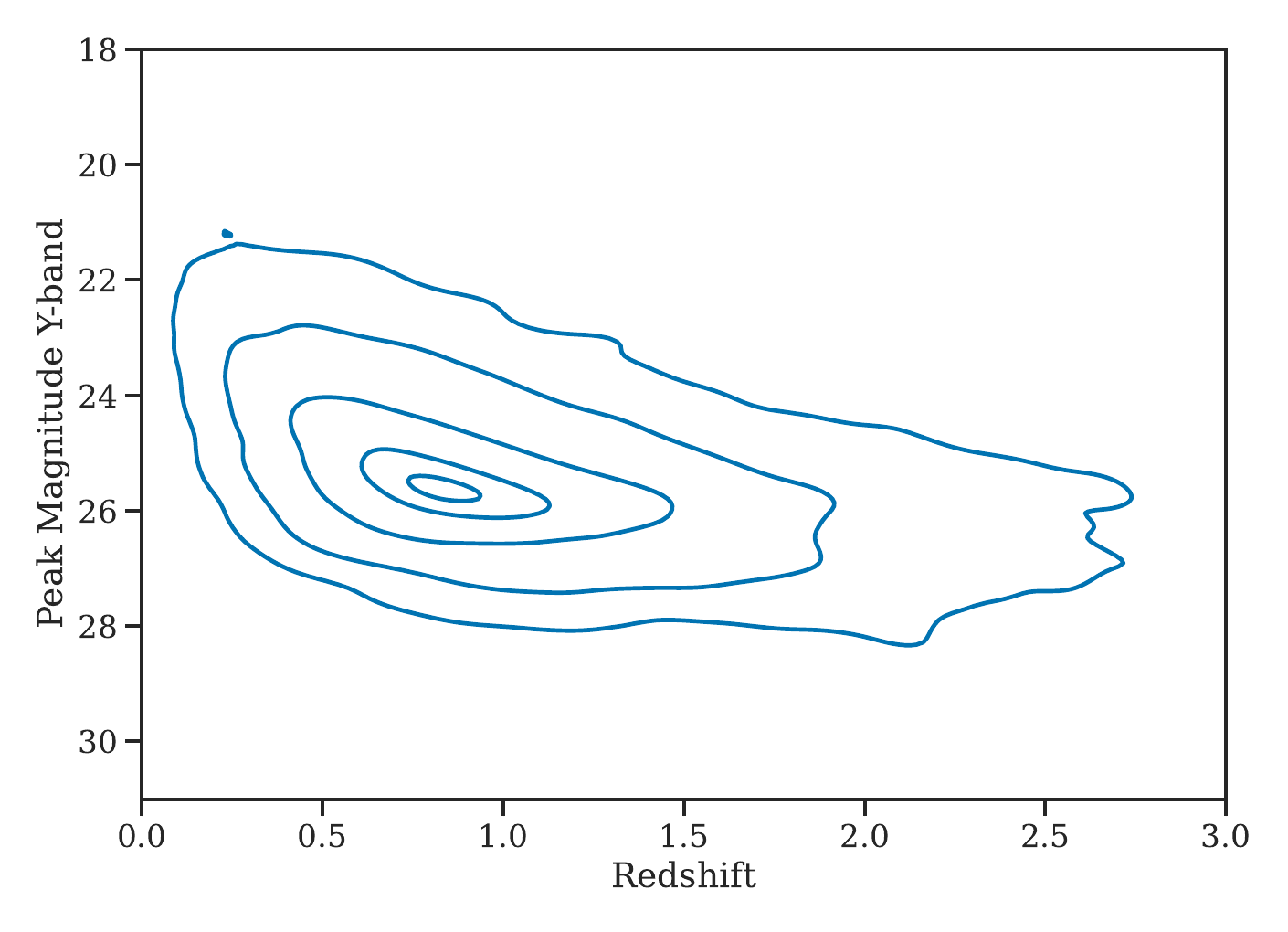}{0.33\textwidth}{CCSN}
               \fig{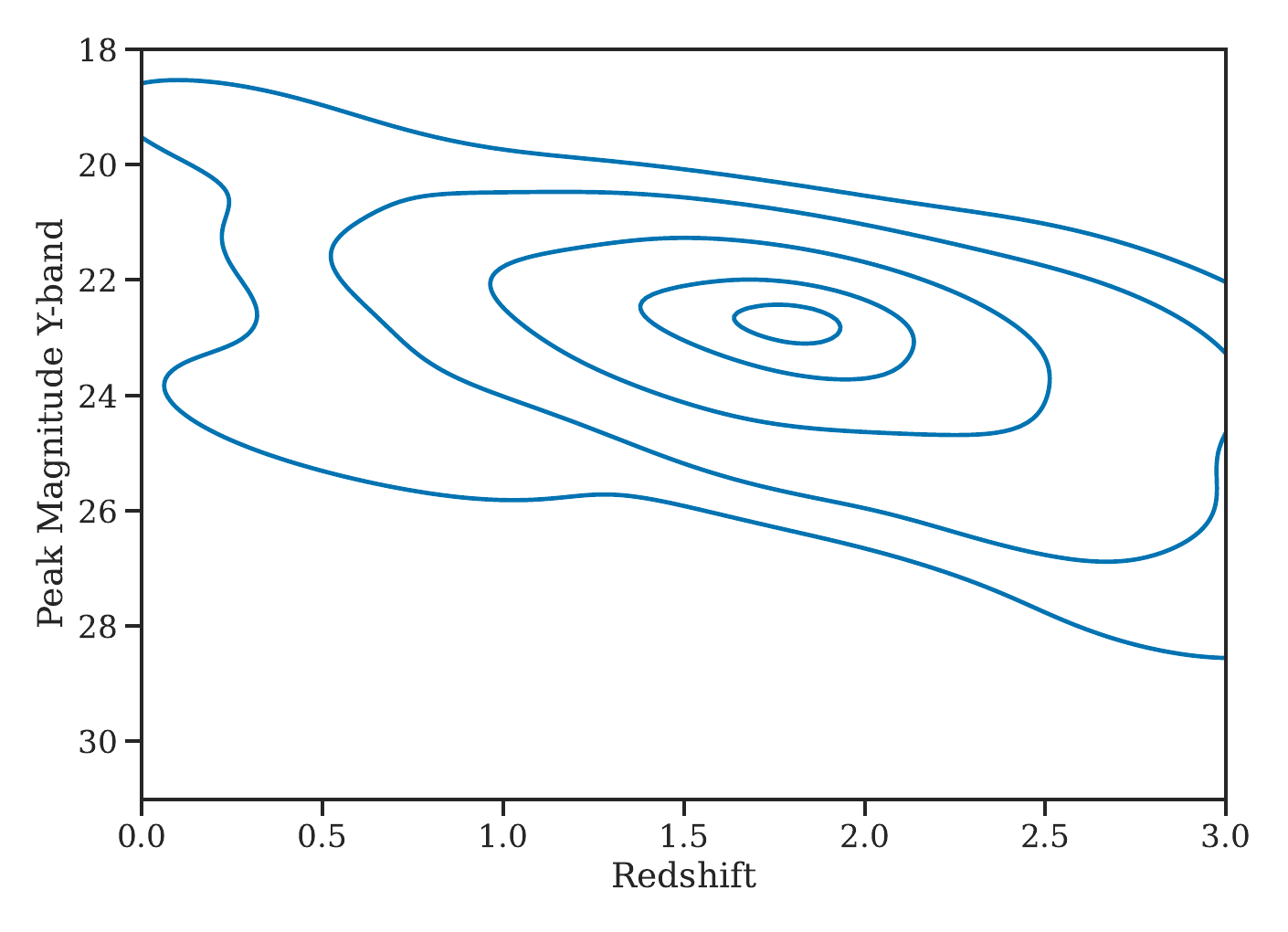}{0.33\textwidth}{SLSN-I}
               \fig{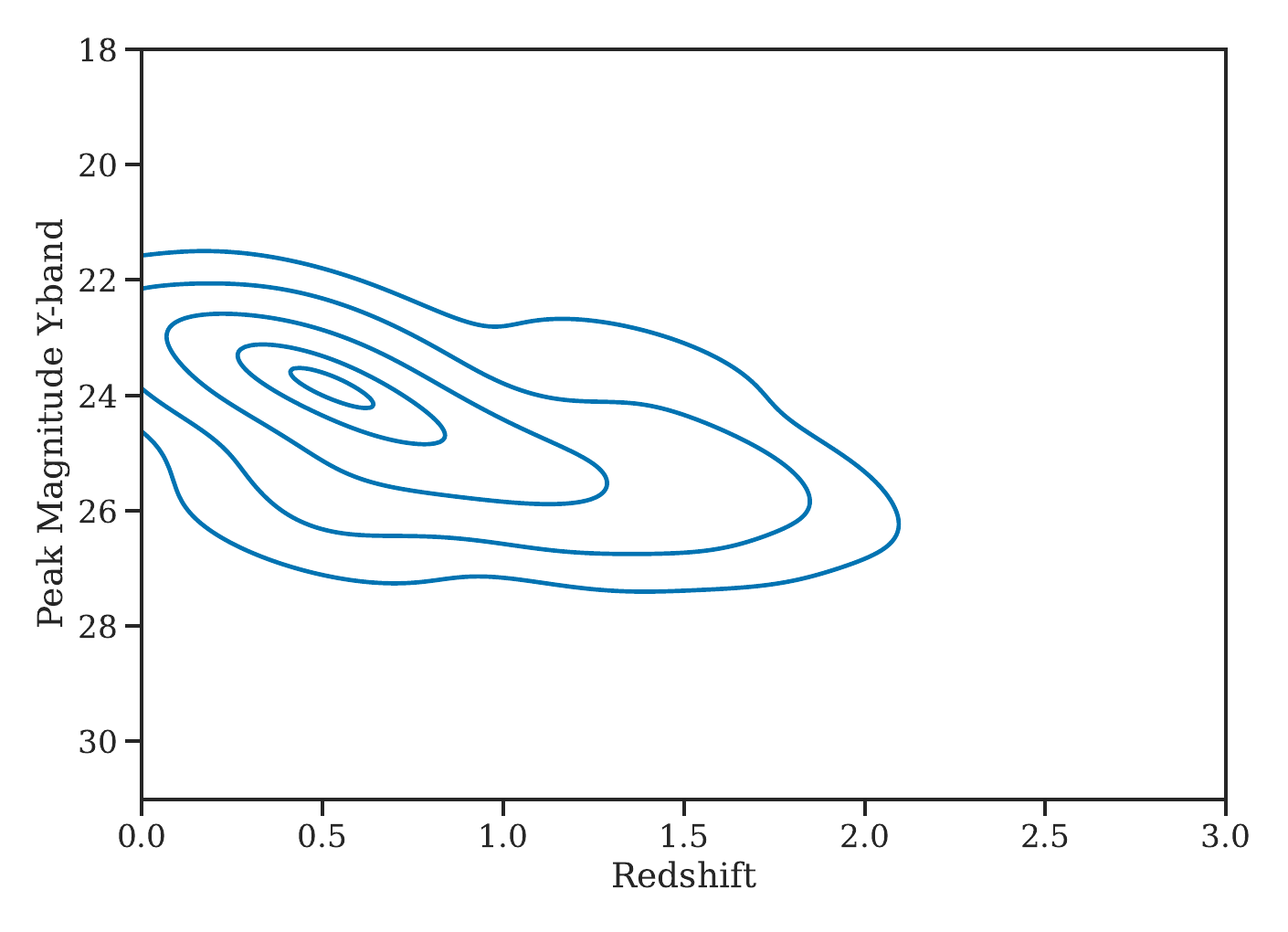}{0.33\textwidth}{TDE}
              }
    \gridline{
               \fig{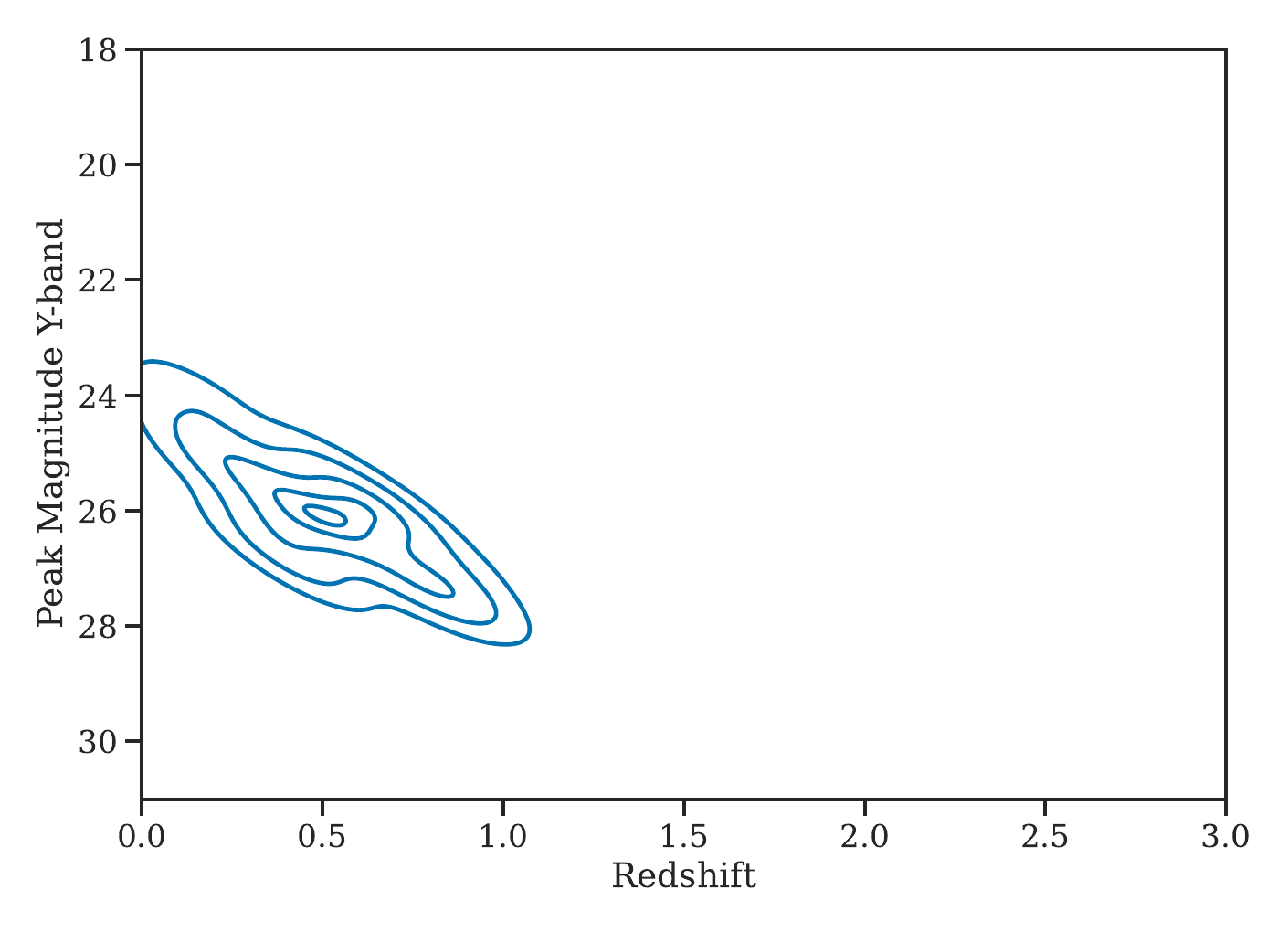}{0.33\textwidth}{ILOT}
               \fig{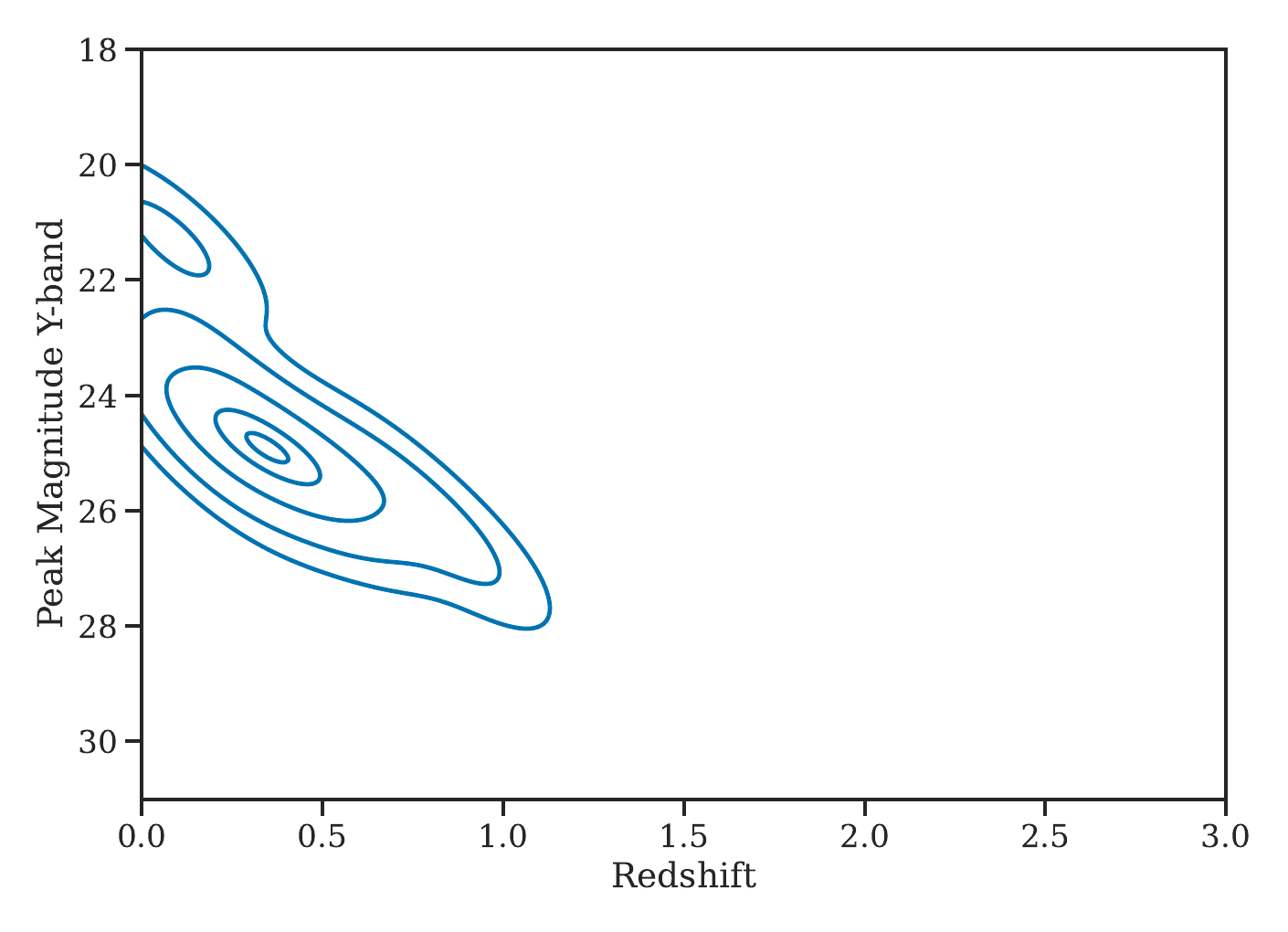}{0.33\textwidth}{KN}
               \fig{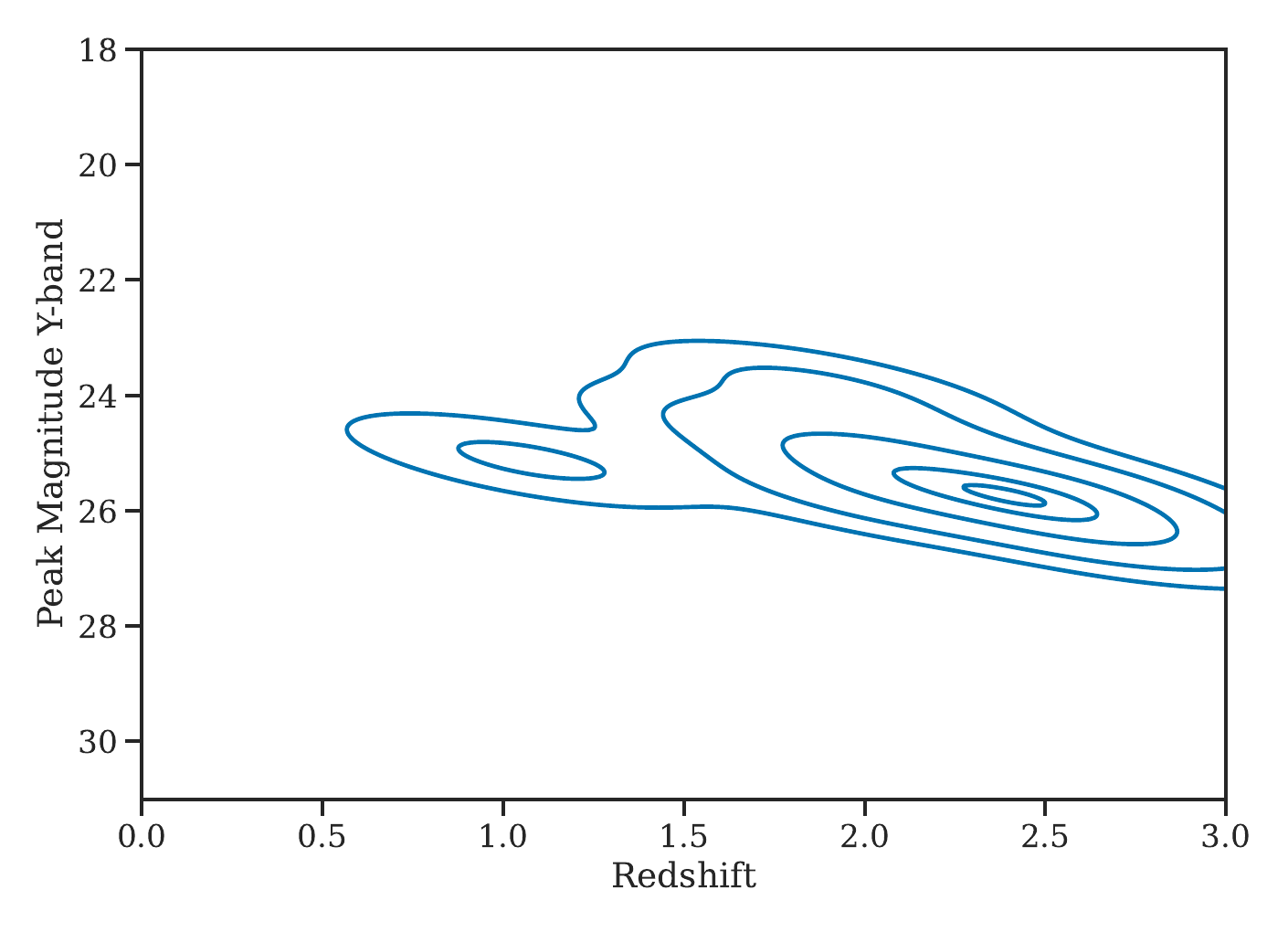}{0.33\textwidth}{PISN}
               }
    \gridline{
               \fig{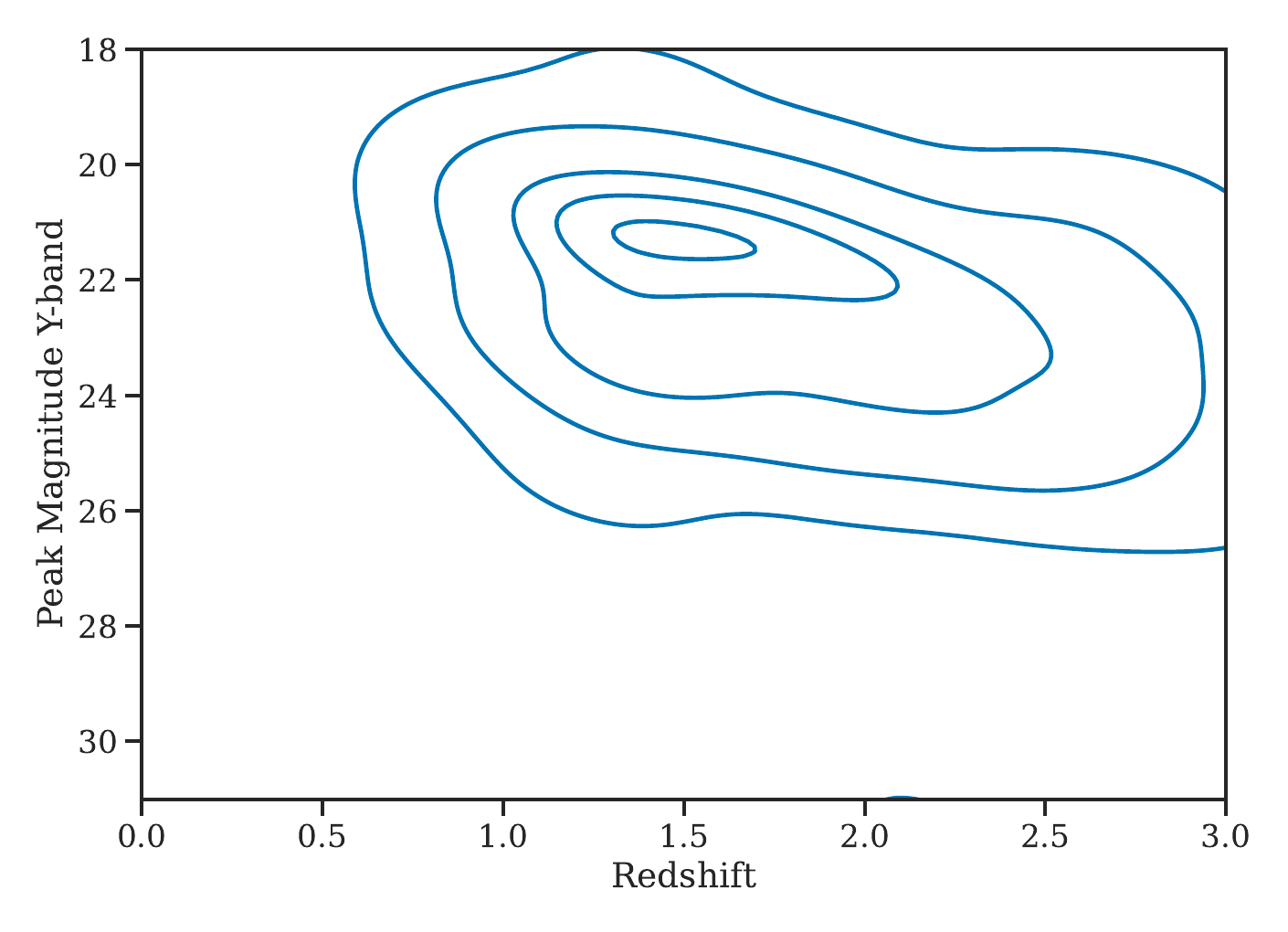}{0.33\textwidth}{AGN}
               }
    \caption{Peak AB-magnitude (in \rst $Y$-band) versus redshift for all \ntrans time-domain classes. Contours are at 2.75, 15.87, 50, 84.13, and 97.25 percentiles. The standard-candle nature of SN~Ia can be seen in the first figure. Additionally, the extreme brightness of SLSN-I is visible in the middle frame. From this plot, it is clear that SLSN-I will be visible at redshifts of $z>3$. 
    There is an artificial cut off at $z=0.08$ for models with a red-edge of 11,000~\AA\xspace since they will not have a defined $Y$-band magnitude.
    }
    \label{fig:mag}
\end{figure*}

To better understand the redshift histograms, we show the peak AB-magnitude in $Y$-band of each object type versus redshift in \cref{fig:mag}.  The precision of SNe~Ia as standard candles is immediately obvious, as they have the tightest scatter in magnitude versus redshift space.  We can also see that the magnitude limits are around 28$^{\mathrm{th}}$~mag.

\begin{figure*}
    \centering
    \includegraphics[width=0.49\textwidth]{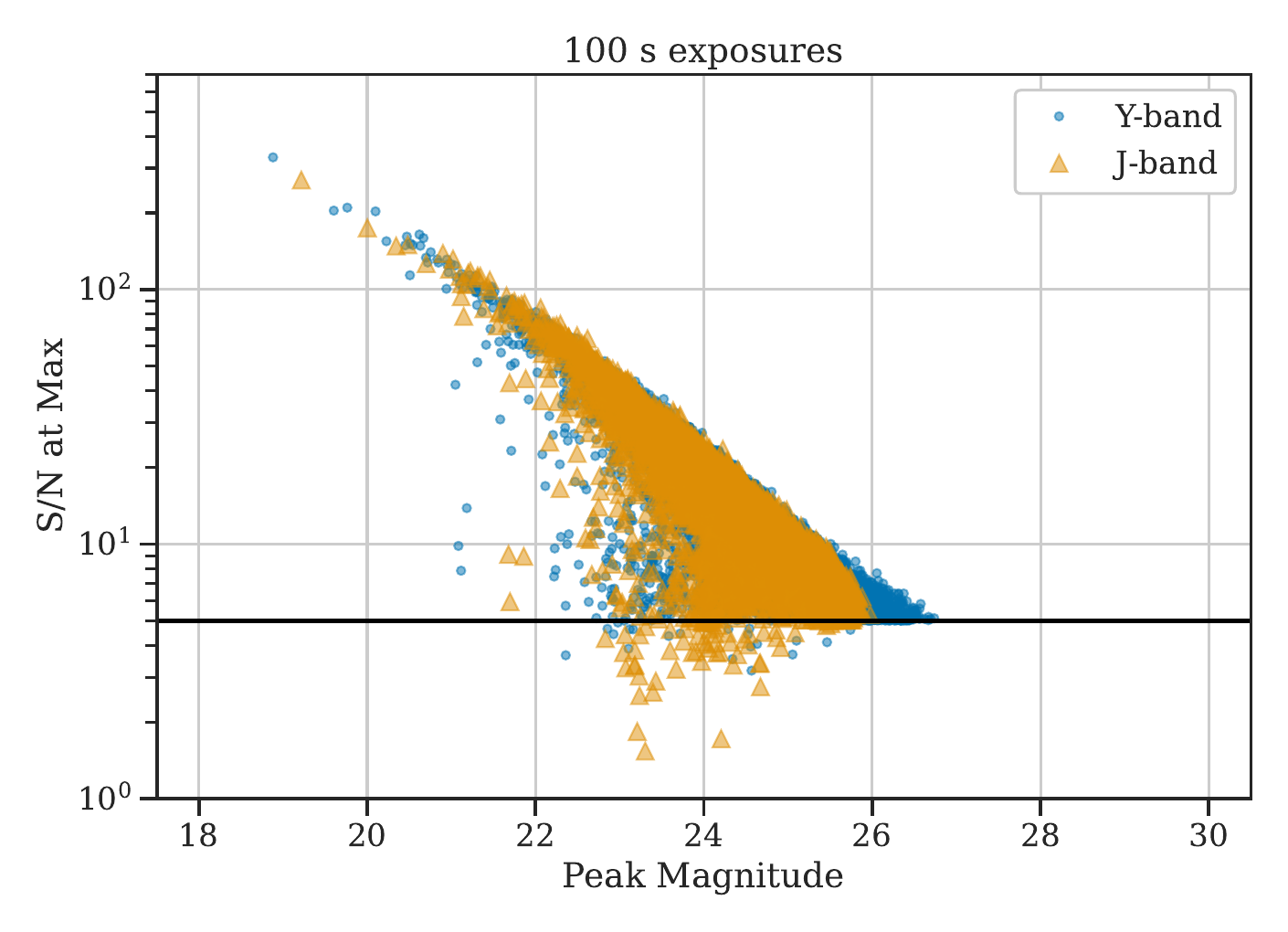}
    \hfill
    \includegraphics[width=0.49\textwidth]{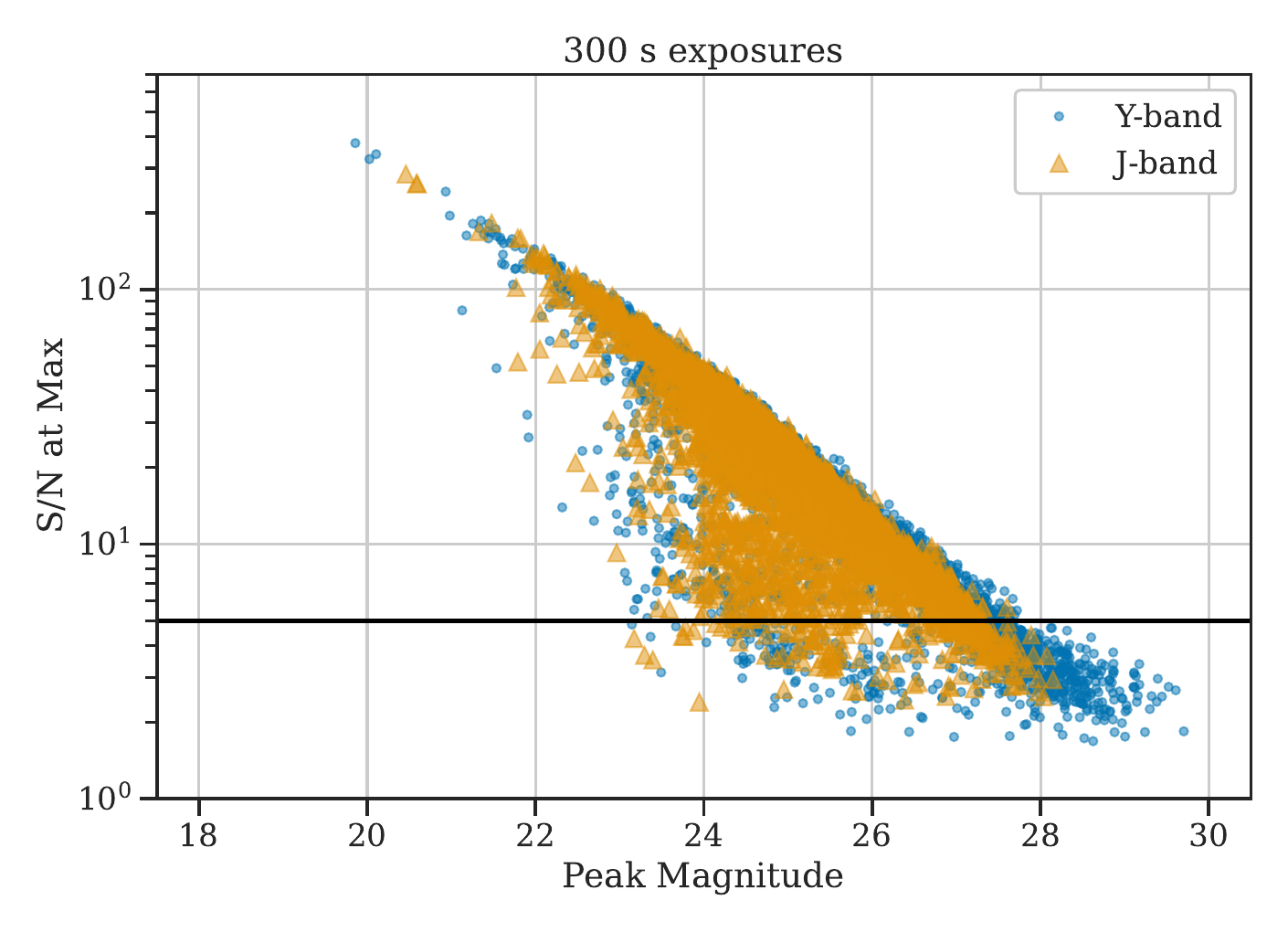}
    \caption{S/N vs.~AB-magnitude at peak luminosity for SNe~Ia in both Y- and J-bands. S/N~=~5 is marked with a horizontal line. The \textit{left} panel shows values taken from the wide tier where Y- and J-band each have 100~s exposures. The \textit{right} panel shows data taken from the deep tier where each band has 300~s exposures. With this exposure time, we calculate a S/N~$\approx 20$ for a SN~Ia of peak magnitude $\sim$25~mag.
    }
    \label{fig:snr}
\end{figure*}

To demonstrate the detection limits per tier, we show the relation between S/N and peak magnitude in \cref{fig:snr}. We find detection limits of S/N of 5 for magnitudes of $\sim$27.5$^{\mathrm{th}}$~mag and $\sim$26.5$^{\mathrm{th}}$~mag for $Y$ and $J$ for deep and shallow respectively.  We are therefore able to translate the relation seen in \cref{fig:mag} to the redshift limits in \cref{tab:results}.

\added{All of these results are highly dependent on the assumed rates. For the most rare transients, such as PISN and KN, the rates can have an uncertainty range of a factor of 2 or 3 (\citet{Briel2022,Pan2012} and \citet{Abbott2021}, receptively). Additionally, rates are poorly defined at the higher redshifts. For example, the \sn rates have only been measured to $z \sim 2$ and above $z > 1$ the uncertainty is significant \citep{Strolger2020}. Yet for this work, we assume these rates are valid to $z = 3$. \romanST's field of view, survey length, and depth will allow it to probe parameter space that has never previously been explored. Though there are high-redshift transits discovered with JWST, \cite[e.g.,][]{Pierel2024a,Pierel2025,DeCoursey2025}, these are not sufficiently large enough samples to improve our rate estimates.}

\begin{figure*}
    \gridline{
              \fig{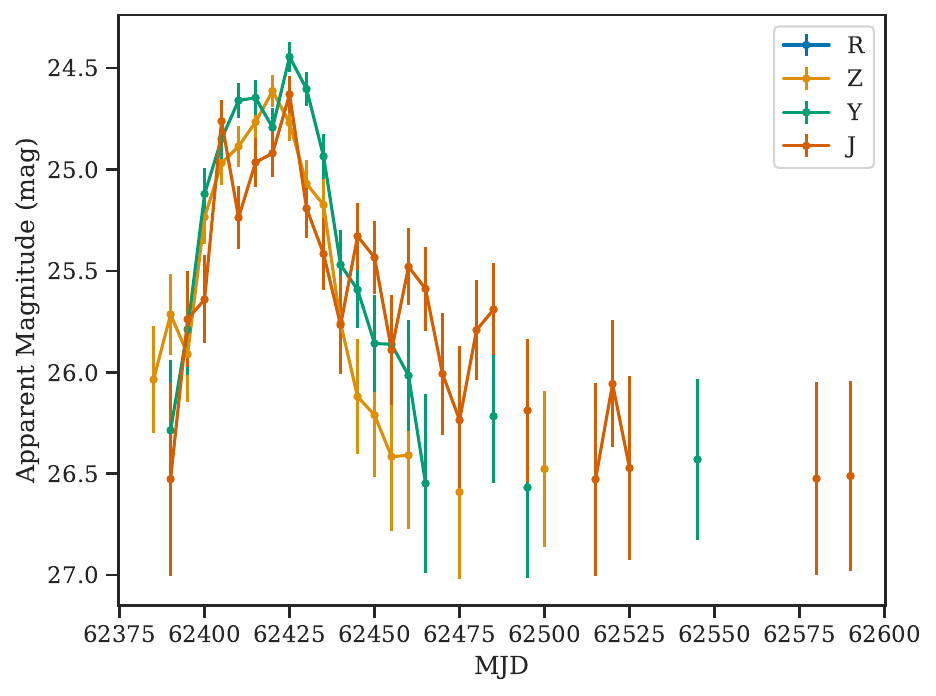}{0.33\textwidth}{SN Ia at $z=1.3$}
              \fig{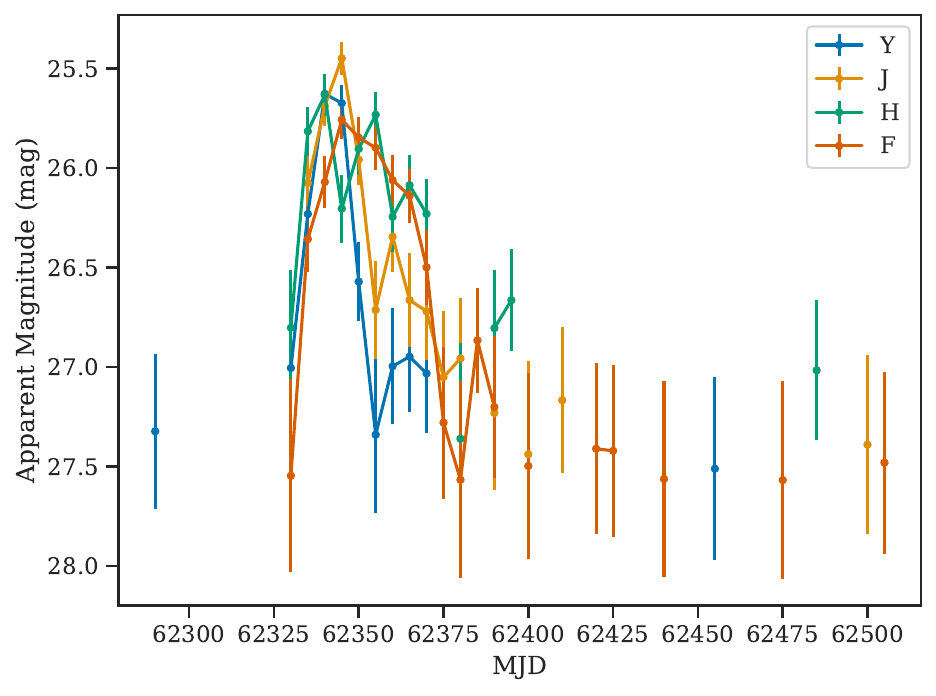}{0.33\textwidth}{SNIa-91bg at $z=1.3$}
              \fig{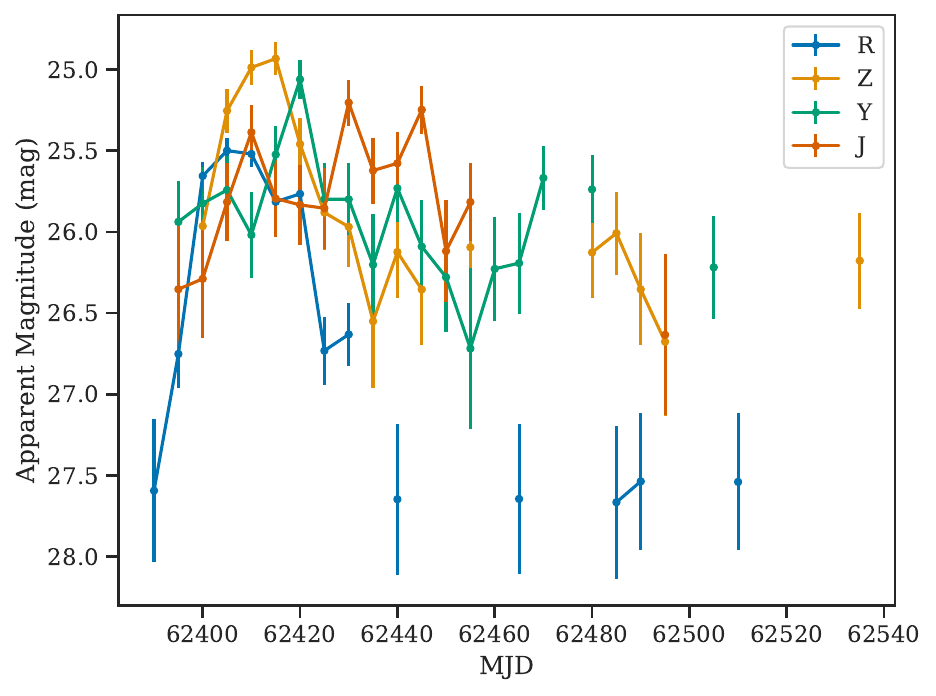}{0.33\textwidth}{SN~Iax at $z=0.4$}
              }
     \gridline{
               \fig{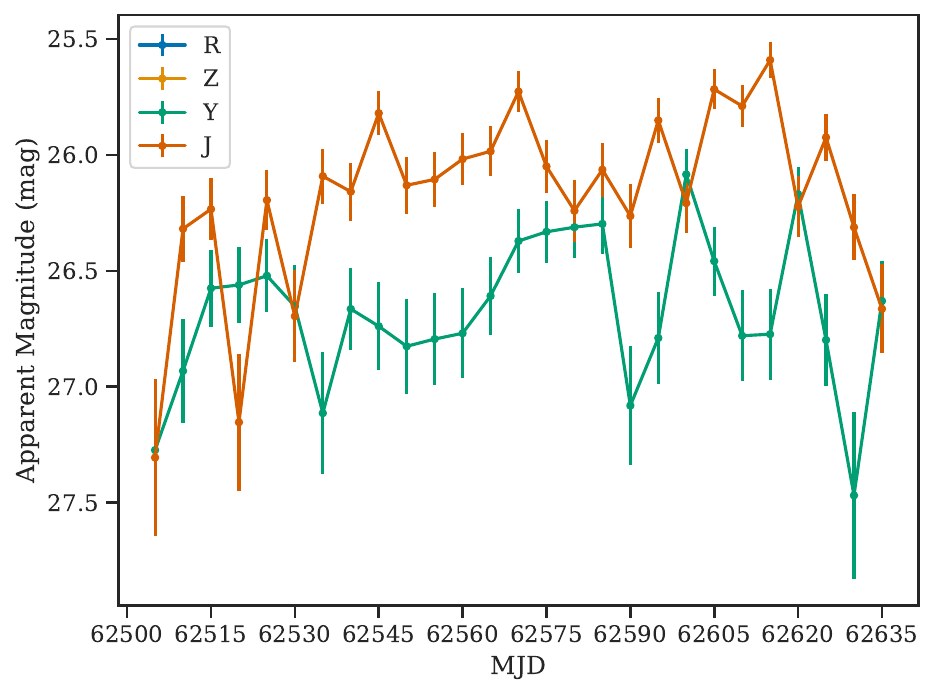}{0.33\textwidth}{CCSN at $z=0.5$}
               \fig{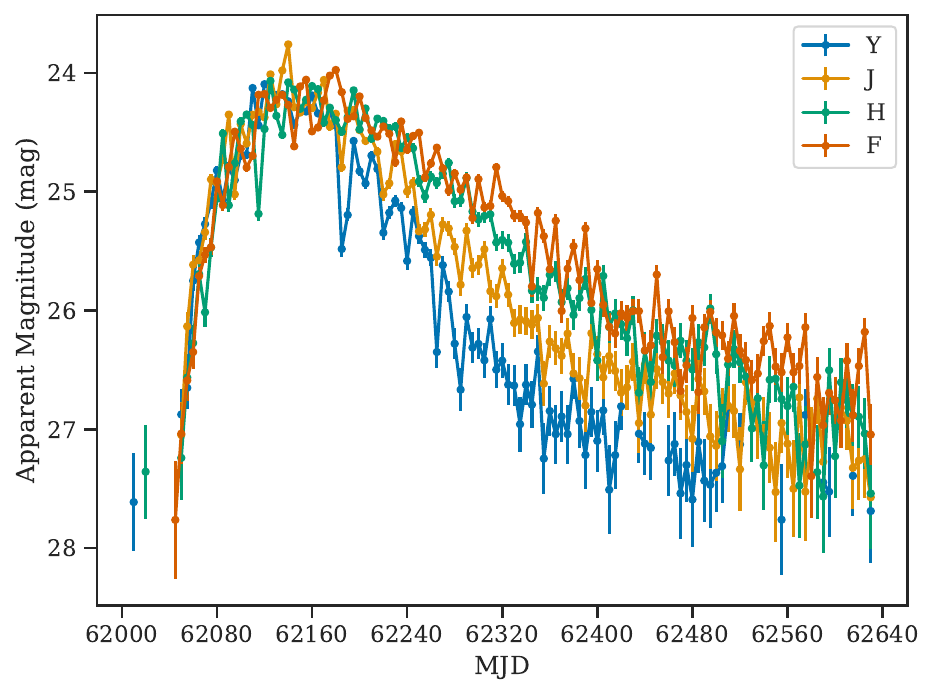}{0.33\textwidth}{SLSN-I at $z=2.4$}
               \fig{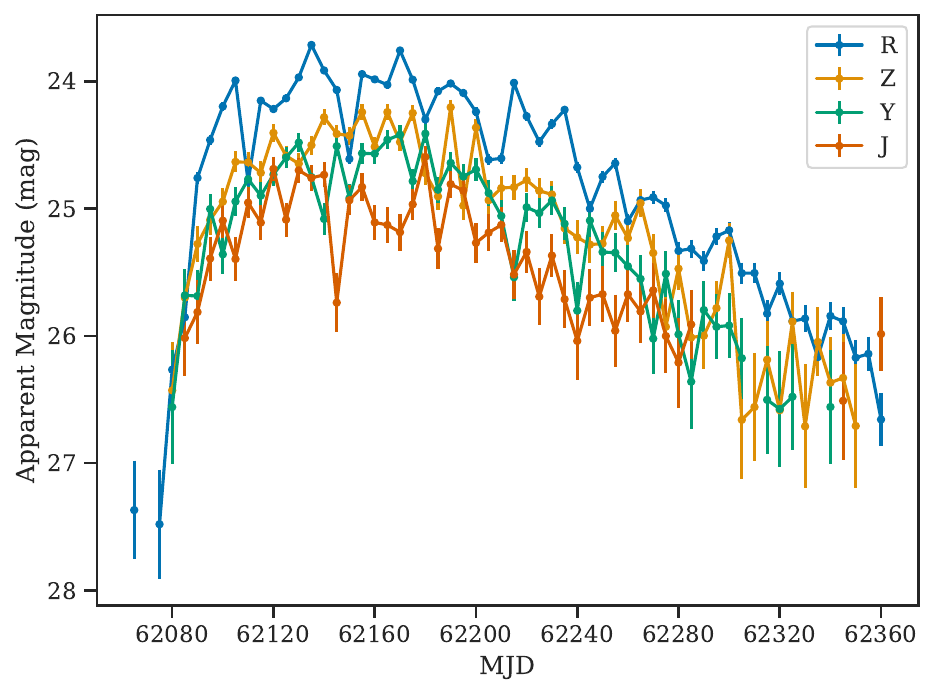}{0.33\textwidth}{TDE at $z=1.4$}
              }
    \gridline{
               \fig{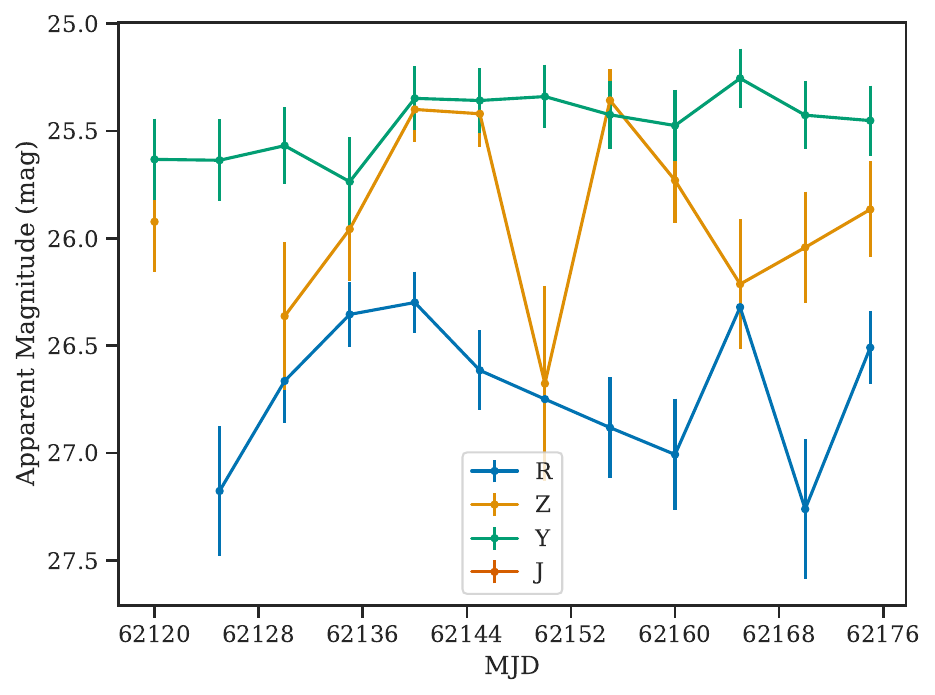}{0.33\textwidth}{ILOT at $z=0.3$}
               \fig{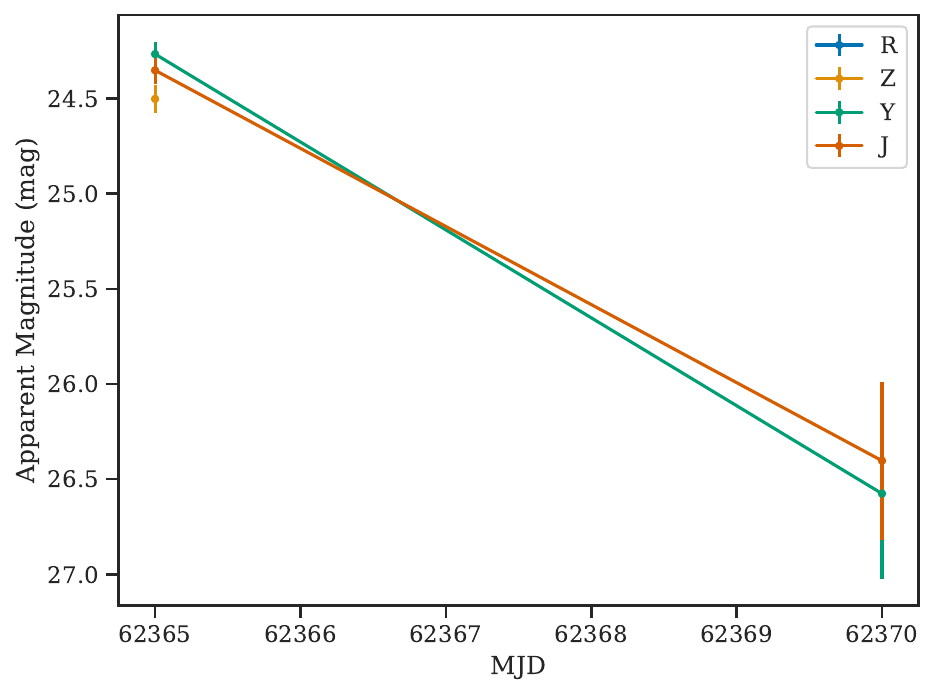}{0.33\textwidth}{KN at $z=0.3$}
               \fig{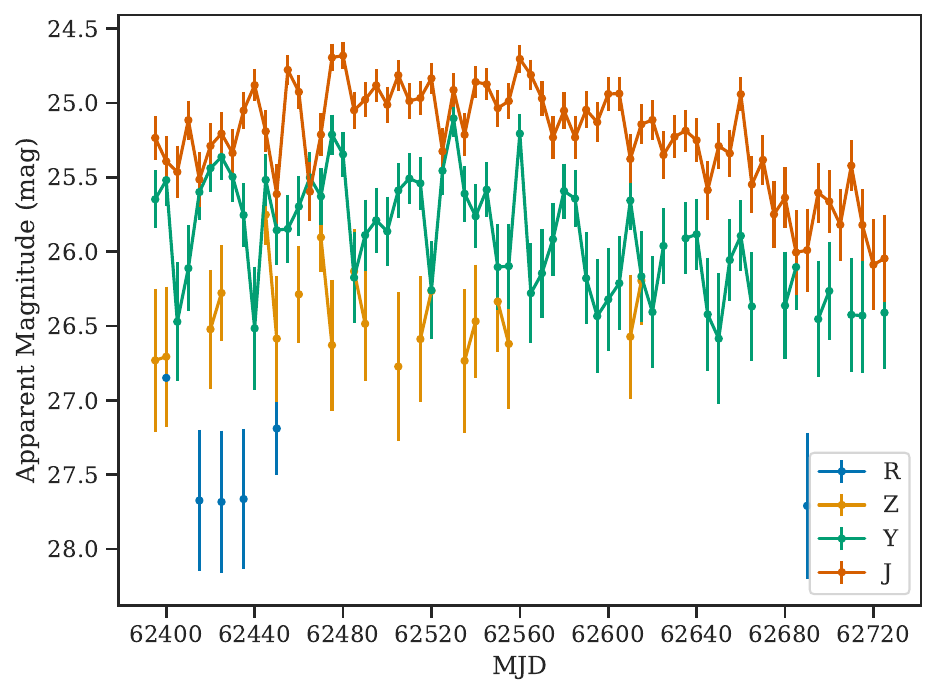}{0.33\textwidth}{PISN at $z=2.6$}
              }
    \gridline{
               \fig{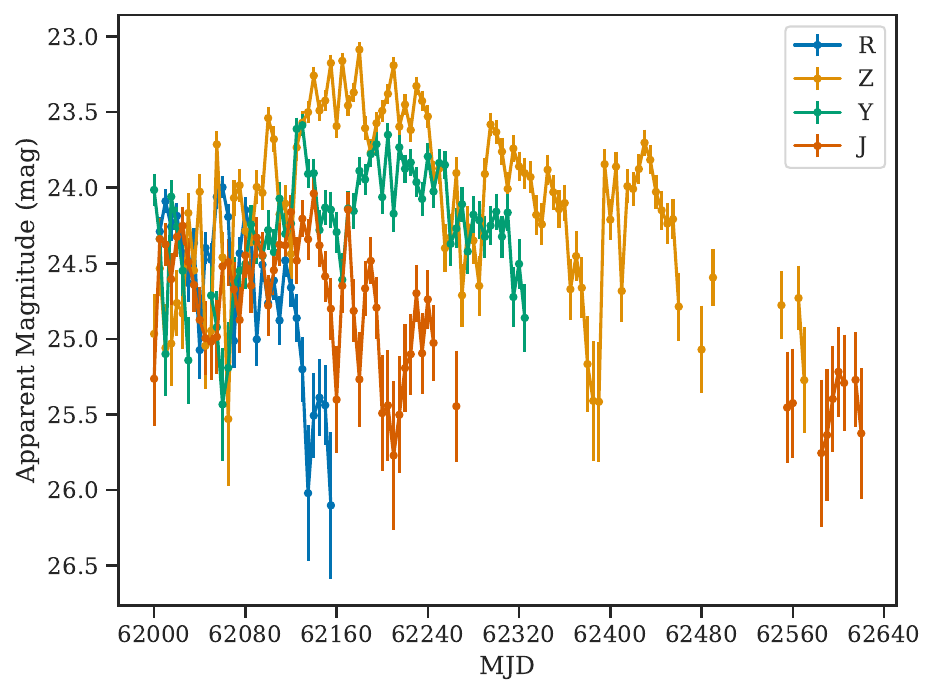}{0.33\textwidth}{AGN at $z=2.8$}
              }
    \caption{Light curves of the median S/N events per type. The long-lived transients, SLSNe and PISNe, have very well sampled light curves. Fast transients are also well sampled, however, the KN detections have measurements around peak, but very little light-curve information. We needed to go to the 75\% percentile S/N at peak in order to get an observation at a second epoch. 
    }
    \label{fig:lc}
\end{figure*}

We showcase example light curves of the median S/N at max for each time-domain class in \cref{fig:lc}. The 5-day observer frame cadence will allow for very high sampling, particularly for the longer lasting transients like SLSNe, TDEs, and PISNe. For KNe, this cadence is closer to the decay rate of the explosions and thus limits their detectability.

\vspace{2em}
\subsection{Comparison with Current Samples}

\begin{figure}
    \centering
    \includegraphics[width=0.95\columnwidth]{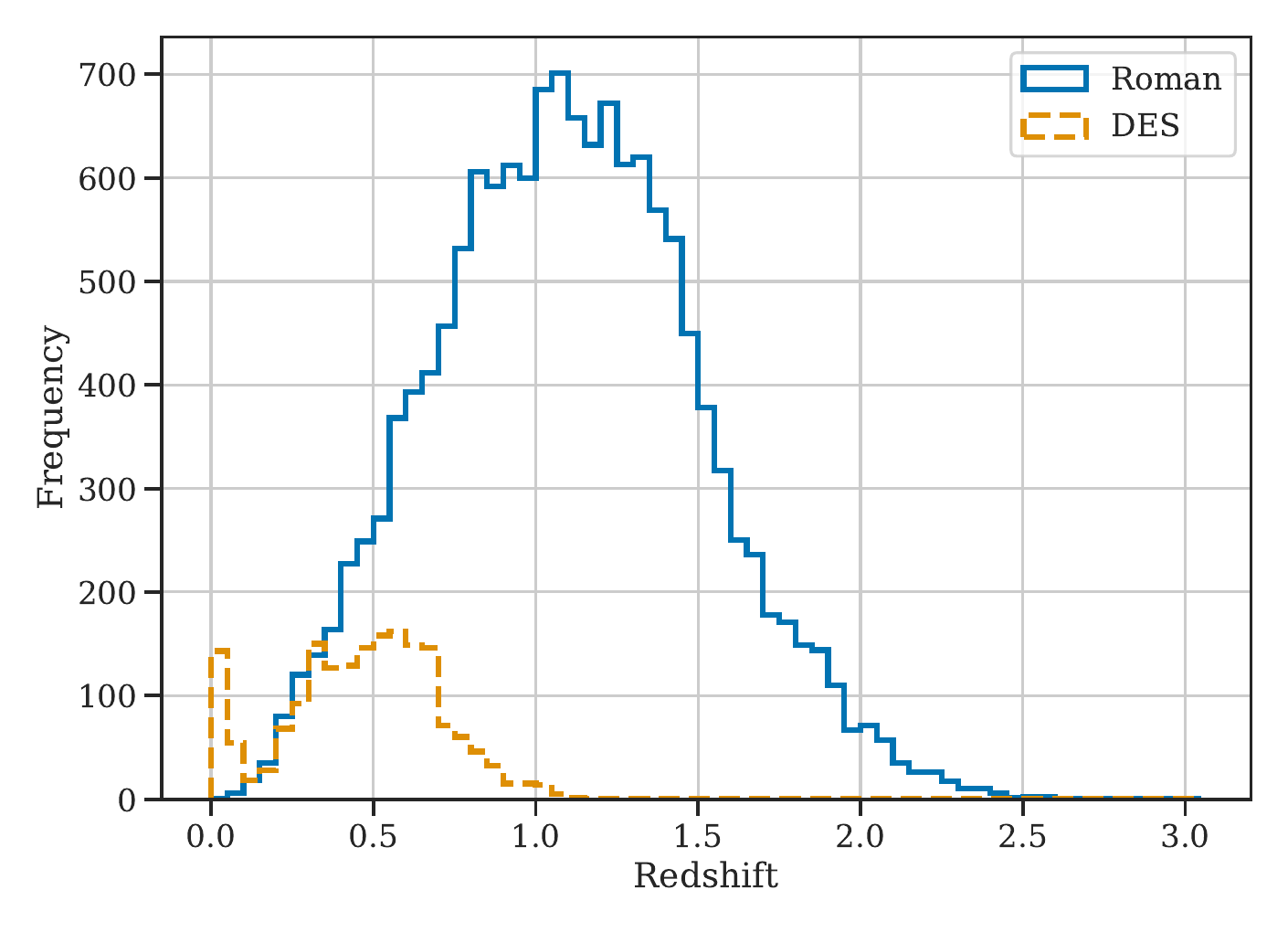}
    \caption{The \rst SN~Ia sample compared to the recent cosmology sample of the Dark Energy Survey. DES has over 1,500 supernovae in its cosmological sameple with very few at $z>1$. However, we expect \rst to have nearly 19,000 SN~Ia with the majority above $z>1$. For this paper, we define ``cosmologically useful'' as having a S/N at max of greater than 10. This is an overly simple assumption, but still reasonable.
    }
    \label{fig:roman-cosmo}
\end{figure}

For SNe~Ia, \romanST will provide an order of magnitude more and reach a redshift of more than twice that of previous cosmological data sets \citep[e.g.,][]{Scolnic2022,Rubin2023,Sanchez2024}. In \cref{fig:roman-cosmo} we compare the redshift distribution of the \rst SN~Ia sample with the Dark Energy Survey Year 5 data release \citep[DES,][]{Sanchez2024}. This is the most recent, photometrically classified sample, and represents the current state of the art for Supernova cosmology.
However, their sample of approximately 1,500 SN~Ia has very few objects above $z>1$. Comparably, \rst is expecting a cosmological sample of over 10,000 SN~Ia. In this simulation, we do not perform a full cosmological analysis so we use a S/N at max of greater than 10 as a proxy for ``cosmological useful''. With this definition, we have over 19,000 Type Ia supernovae with the majority being above a redshift of one.

\subsection{Spectra}

\begin{figure}
    \centering
     \includegraphics[width=0.45\textwidth]{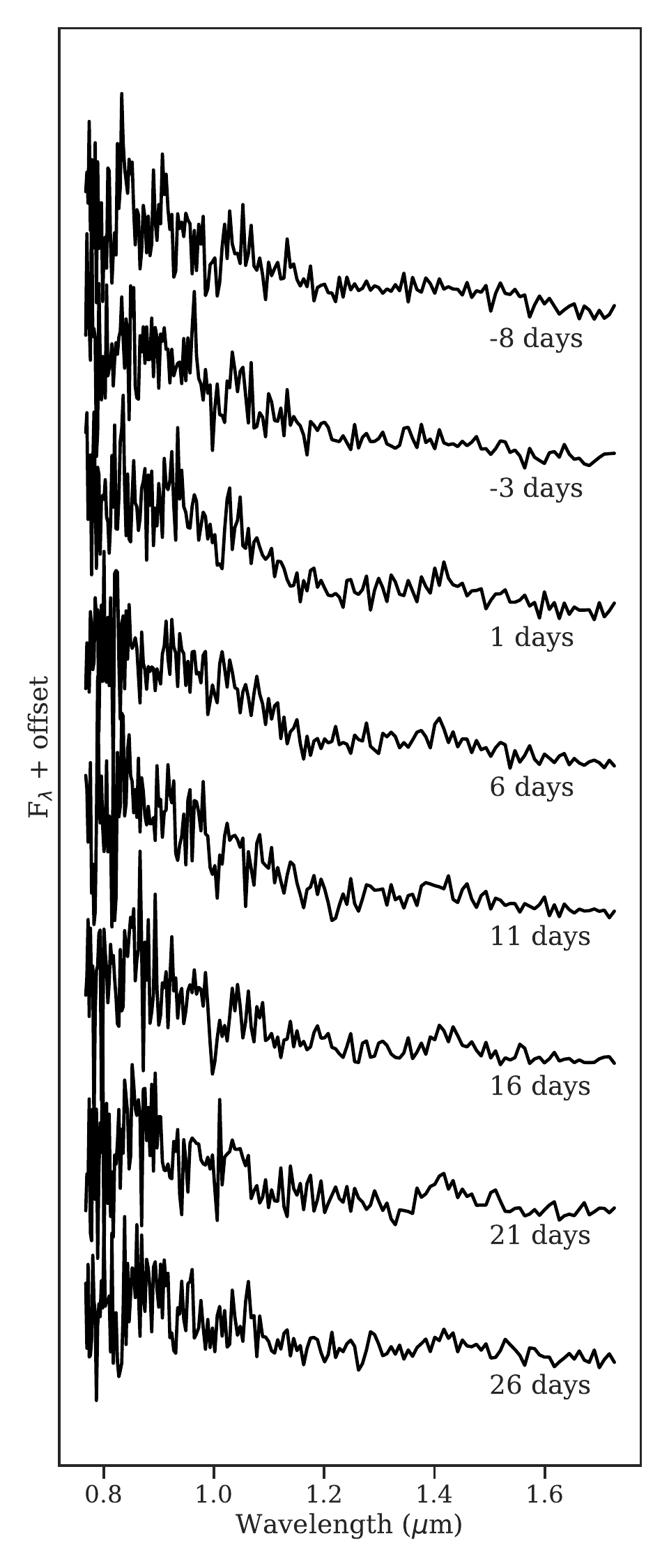}
    \caption{A spectral time series from the \rst WFI Prism, presented in the observer's frame. This is a \replaced{$z=0.579$}{$z=0.636$} SN~Ia observed in the wide tier (900~s exposures). This results in 9 observations from phase \replaced{-6 to 28}{-8 to 26} days.
    }
    \label{fig:spec}
\end{figure}

In addition to the photometry, we simulate the prism spectral time series for all \ntrans time-domain classes. As an example, we present one SN~Ia as observed with the wide-tier exposures, 900~s, in \cref{fig:spec}. We show \replaced{9}{8} phases, from phase \replaced{-9 to 30}{-6 to 28} days. Additional late time spectra will be observed with \rst, but our SN~Ia model stops at a phase of 40 days.

\subsection{Photometric Classification of Type Ia Supernovae}

For a first test on these simulations, we train and test the binary classification of SCONE \citep{Qu2021}. SCONE is a convolutional neural network that was used in the DES SN cosmological analysis. For training we use a unique set of only 8,600 SNIa and 8,600 ``contamination'' objects (CCSN, SN~Iax, SNIa-91bg). 
This simulation uses the same properties as the main Hourglass simulations. The resulting SCONE model can be found at \url{https://github.com/Roman-Supernova-PIT/hourglass_snana_sims/tree/main/3_clas/models/scone}.

\begin{figure}
    \centering
    \includegraphics[width=0.95\columnwidth]{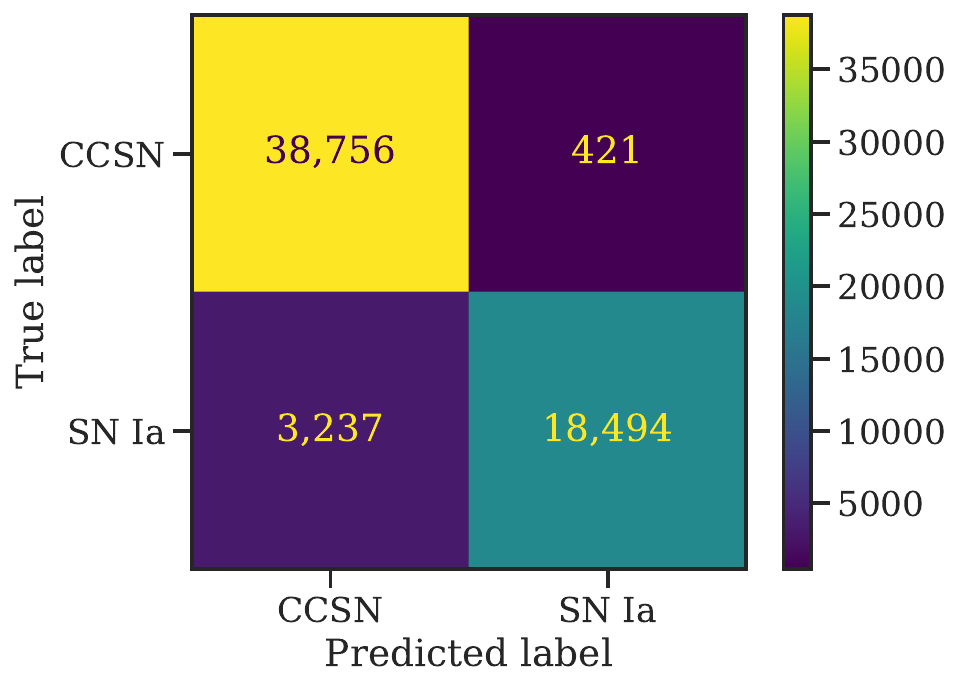}
    \caption{Confusion matrix for the photometric classifier SCONE. We train the binary classification to distinguish between cosmologically useful SNe~Ia and contaminates. Contaminates are labeled as CCSN but also contains SNIa-91bg and SN~Iax objects. SCONE gets a high recall (85\%, i.e. incorrectly classifying a small fraction of the SNe~Ia). More importantly, we obtain a high precision (\replaced{96}{98}\%), meaning that the predicted SNe~Ia label has a low level of contamination.
    }
    \label{fig:confusion}
\end{figure}

SCONE requires more than just a single detection, therefore, we add a requirement that there are minimum 5 epochs per transient. This results in 21,731 SNe~Ia and \replaced{47,040}{39,177} combined CCSNe, SN~Iax, and SNIa-91bg objects.
The confusion matrix can be seen in \cref{fig:confusion}. We obtain an accuracy of 94\% (number of correct predictions over all predictions), along with a precision of \replaced{96}{98}\% (correctly classified SN~Ia over everything classified as a SN~Ia) and a recall of 85\% (correctly classified SN~Ia over everything that \textit{is} a SN~Ia). 

\begin{figure}
    \centering
    \includegraphics[width=0.95\columnwidth]{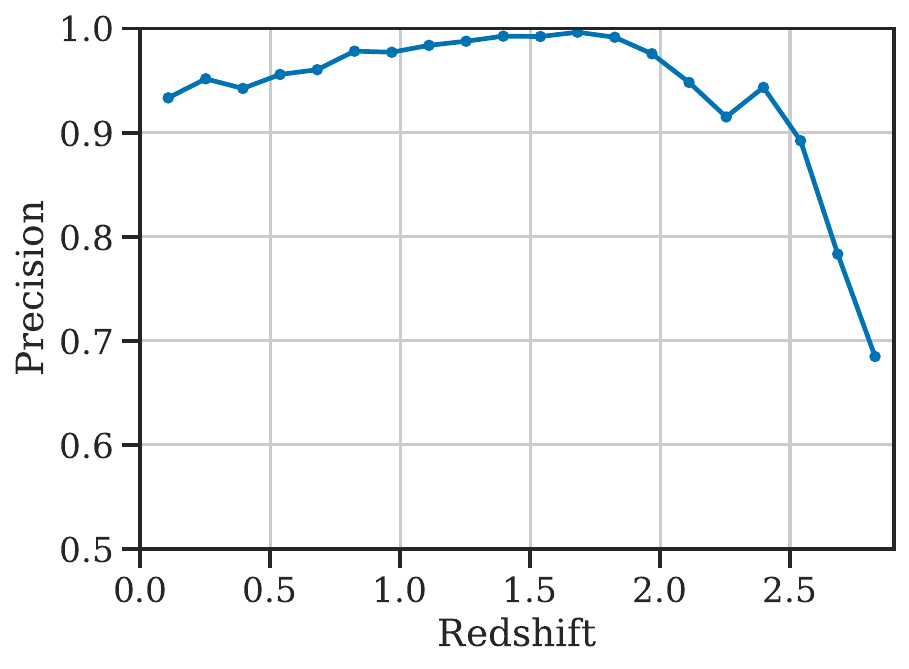}
    \caption{SCONES's precision is around \replaced{95}{99}\% and peaks \replaced{between redshifts of 1.5 and 2.0}{near a redshift of 1.7}. However, it has a significant drop off at high redshifts. The slightly worse performance at low-z is unexpected. We are unsure if this is a issue with low statistics or the fact that the rest-frame near-infrared features of the transients are different.
    }
    \label{fig:precision}
\end{figure}

We investigate SCONE's performance as a function of redshift. We present SCONE's precision, the fraction of correctly labeled SNe~Ia over of all events labeled SNe~Ia, as a function of redshift in \cref{fig:precision}. SCONE predominately performs at $>$95\% but has a significant drop off at high redshifts \added{($z > 2.5$)}.

\section{Conclusions}\label{sec:conclusions}

In this work, we present and release the Hourglass simulation, a catalog for the \romanFull High-Latitude Time-Domain Core Community Survey. This simulation includes \ntrans extra-galactic time-domain classes. We run this through the current baseline \rst survey: four filters per tier, a five day cadence, a wide tier of 19~deg$^2$, and a deep tier of 4.2~deg$^2$, with about $\sim$20\% of the area also covered by prism observations. 

With a detection threshold of S/N $> 5$ in two filters, we find that a \rst time-domain catalog would have approximately \replaced{26,000}{21,700} SNe~Ia and \replaced{70,000}{39,000} CCSNe. There would also be \replaced{over}{around} 70 SLSN, 40 TDEs, over 10 PISNe, and \replaced{around 5}{possibly 3} KN. We demonstrate that the data set is not purely low S/N observations: most transient classes have a median S/N at maximum luminosity of $\sim$10, and long-lived events have individual light curves with 200 to 500 observations. Additionally, we present the first examples of the expected spectral-time series data from \rst's prism and the SCONE photometric classifier on \rst data. The simulated data can be found at \doi{10.5281/zenodo.14262943}.

Though this simulation surpasses what has come before---in terms of breadth of transient types and the inclusion of the spectral time series data---it is still insufficient for the community's preparation for the \romanST. Future simulations need to take into account the hardware knowledge gained from thermal-vacuum testing, test observation \replaced{strategies beyond the reference survey}{strategy recommended by the ROTAC}, and continue to add time domain sources such as variable stars. Currently, we don't expect these changes to have a significant effect on what the \rst High-Latitude Time-Domain Core Community Survey catalog looks like\added{, other than increasing the number of transients detected}, but we have set up the Hourglass simulation in such a way that we can update it when new input parameters are available.

\begin{acknowledgments}
The authors thank Qifeng Cheng and Konstantin Malanchev for access to the AGN model used in ELAsTiCc.

Funding for the \rst Supernova Project Infrastructure Team has been provided by NASA under contract to 80NSSC24M0023
This work was completed, in part, with resources provided by the University of Chicago’s Research Computing Center.
M.V.\ was partly supported by NASA through the NASA Hubble Fellowship grant HST-HF2-51546.001-A awarded by the Space Telescope Science Institute, which is operated by the Association of Universities for Research in Astronomy, Incorporated, under NASA contract NAS5-26555. 
\added{Work by R.H.\ and this material is based upon work supported by NASA under award number 80GSFC24M0006.}
D.S. is supported by DOE grant DE-SC0010007, DE-SC0021962, the David and Lucile Packard Foundation, and the John Templeton Foundation.
R.K. is supported by DOE grant DE-SC0009924.

B.R. lead the project, ran the sims and wrote the first draft of the paper.
M.V. assisted in running and analyzing the non-SN~Ia the simulations.
R.H. provided guidance and help in ensuring the \rst specific components were accurately simulated and provided details from the first set of \rst simulations using \texttt{SNANA}.
H.Q. supported the development and use of the photometric classifier SCONE.
L.A. helped validate the data release files.
D.S. advised and worked on figures.
R.K. supported the development and debugging of \texttt{SNANA}.
P.M. helped get the project started with insight from previous \rst simulations using \texttt{SNANA}.
D.B. provided a technical review of the paper.
All authors provided editing and discussions around the paper and the data release.
\end{acknowledgments}

\software{Matplotlib \citep{matplotlib}, 
Numpy \citep{numpy2020}, 
Pandas \citep{pandas}, 
PIPPIN \citep{Hinton2020},
Python, 
SciPy \citep{scipy}, 
SCONE \citep{Qu2021},
Seaborn \citep{seaborn},
\texttt{SNANA} \citep{Kessler2009a,Kessler2017}.
}

\bibliographystyle{apj}
\bibliography{library}

\appendix
\section{Data Release}\label{data_release}

The data is available at \doi{10.5281/zenodo.14262943}. It comes in three parquet files: \texttt{hourglass\_objects.parquet}, \texttt{hourglass\_photometry.parquet}, and \texttt{hourglass\_spectra.parquet}.
In python, these files can be read with Pandas via the \texttt{.read\_parquet()} method. PyArrow also has a \texttt{.read\_parquet()} method. PyArrow's method allows for the inspection of the file prior to importing the whole thing into memory.
One useful thing is to inspect the metadata with
\texttt{pq.ParquetFile(`hourglass\_objects.parquet').metadata.metadata}. This provides information like the version of the data set and the date it was produced.

The object file is one row per object. It has all the ``object-level'' information like RA, Decl, redshift, S/N at maximum, and other information. The full list of columns for this file can be seen in \cref{tab:obj_table}. \added{For operations, we expect these data releases to be richer. We release this set of simulations to engage the community to determine what light curve features are most useful to include.}

The next file is the photometry file. It has one row per flux measurement. Each row has the ``CID", for joining with the ``CID" column in the object file. Each row also has the flux, band, PSF, sky noise, and other photometry measurements. The full list can be seen in \cref{tab:phot_table}.

Finally we provide the spectroscopic data. This file is one row per object per epoch. Each row has a ``CID'' for joining with the object table. The full list of columns in this table can be found in \cref{tab:spec_table}.

\begin{deluxetable*}{llcr}
% \tablecolumns{2}
% \tabletypesize{\small}
\tablewidth{0pt}
\tablecaption{Description of \texttt{hourglass\_objects.parquet}\label{tab:obj_table}}
\tablehead{ 
    \colhead{Column Name} & \colhead{Description Name} & \colhead{Units} & \colhead{Type}
}
\startdata
  cid & Candidate ID & \nodata & integer \\
  \hline
  field & Field name & \nodata & string\\
  class & Object classification name & \nodata & string\\
  sub\_class & Name of subclass, such as ``IIP" or ``Ic" & \nodata & string \\ %SIM_TYPE_NAME
  z\_cmb & Redshift in CMB frame & \nodata & float\\
  ra & Right Ascension & degrees & float \\
  dec & Declination & degrees & float\\
  mw\_ebv & Milky-way $E(B-V)$ along ling-of-sight & AB mag & float\\
  \hline
  peak\_mjd & Modified Julian date of maximum & \nodata & float\\
  snr\_max\_Y & Simulated peak S/N in \textit{Y}-band & \nodata & float\\
  snr\_max\_J & Simulated S/N magnitude in \textit{J}-band & \nodata & float\\
  peak\_mag\_Y & Simulated peak magnitude in \textit{Y}-band  & AB mag & float\\
  peak\_mag\_J & Simulated peak magnitude in \textit{J}-band & AB mag & float\\
  \hline
  n\_obs & Number of total observations & \nodata & integer\\
  t\_rest\_min & Phase of first observation & days & float\\
  t\_rest\_max & Phase of last observation & days & float\\
  \hline
  scone\_prob\_Ia & Probability of being a SN Ia via SCONE & \nodata & float
\enddata
% \tablecomments{
% Detected values are rounded according to the expected counting statistics uncertainty.
% }
\end{deluxetable*}

\begin{deluxetable*}{llcr}
% \tablecolumns{2}
% \tabletypesize{\small}
\tablewidth{0pt}
\tablecaption{Description of \texttt{hourglass\_photometry.parquet}\label{tab:phot_table}}
\tablehead{ 
    \colhead{Column Name} & \colhead{Description Name} & \colhead{Units} & \colhead{Type}
}
\startdata
  cid & Candidate ID & \nodata & integer \\
  \hline
  mjd & Modified Julian date of observation & \nodata & float\\
  band & Photometric band, one of R, Z, Y, J, H, or F & \nodata & string \\
  % FIELD & Field name & string\\
  \hline
  phot\_flag & Photometric Flag, 0 is pass & \nodata & integer\\
  fluxcal & Calibrated flux, mag~$= 27.5 - 2.5\log_{10}$(``fluxcal'') & \nodata & float\\
  fluxcal\_err & Poisson uncertainty on ``fluxcal'',  sky+galaxy+source & \nodata & float\\
  \hline
  psf\_nea & PSF noise equivalent area & pixels & float\\
  sky\_sig & Sky noise & ADU/pixel & float\\
  read\_noise & Read noise & ADU/pixel & float\\
  zp & Zero-point & AB mag & float\\
  zp\_err & Error on zero-point & AB mag & float \\
  \hline
  sim\_mag\_obs & Input model mag & AB mag & float\\
\enddata
% \tablecomments{
% Detected values are rounded according to the expected counting statistics uncertainty.
% }
\end{deluxetable*}

% \begin{deluxetable*}{llr}
% % \tablecolumns{2}
% % \tabletypesize{\small}
% \tablewidth{0pt}
% \tablecaption{Description of \texttt{hourglass\_objects.parquet}\label{tab:obj_table}}
% \tablehead{ 
%     \colhead{Column Name} & \colhead{Description Name} & \colhead{Format}
% }
% \startdata
%   CID & Candidate ID & integer \\
%   \hline
%   FIELD & Field name & string\\
%   CLASS & Object classification name & string\\
%   % NON1A\_INDEX &  & ??\\
%   ZCMB & Redshift in CMB frame & float\\
%   RA & Right Ascension, degrees & float \\
%   DEC & Declination, degrees & float\\
%   MWEBV & Milky-way $E(B-V)$ along ling-of-sight & float\\
%   \hline
%   PEAKMJD & Modified Julian date of maximum & float\\
%   SNRMAX\_Y & Simulated peak S/N in \textit{Y}-band & float\\
%   SNRMAX\_J & Simulated S/N magnitude in \textit{J}-band & float\\
%   PEAKMAG\_Y & Simulated peak magnitude in \textit{Y}-band  & float\\
%   PEAKMAG\_J & Simulated peak magnitude in \textit{J}-band & float\\
%   \hline
%   NOBS & Number of total observations & integer\\
%   TRESTMIN & Phase of first observation, days & float\\
%   TRESTMAX & Phase of last observation, days & float\\
%   \hline
%   SCONE\_PROB\_IA & Probability of being a SN Ia via SCONE & float
% \enddata
% % \tablecomments{
% % Detected values are rounded according to the expected counting statistics uncertainty.
% % }
% \end{deluxetable*}

\begin{deluxetable*}{llcr}
\tablecolumns{3}
% \tabletypesize{\small}
% \tablewidth{0pt}
\tablecaption{Description of \texttt{hourglass\_spectra.parquet}\label{tab:spec_table}}
\tablehead{ 
    \colhead{Column Name} & \colhead{Description Name} & \colhead{Units} & \colhead{Type}
}
\startdata
  cid & Candidate ID & \nodata & int \\
  \hline
  mjd & Modified Julian date of observation & \nodata & float \\
  t\_expose & Exposure time & seconds & float\\
  n\_bin\_lam & Number of wavelength bins & \nodata & int\\
  \hline
  lam\_min & Low wavelength edge of spectral bin & angstroms & list of floats \\
  lam\_max & High wavelength edge of spectral bin & angstroms & list of floats\\
  flam & Flux per wavelength bin & erg/s/\AA/cm$^2$ & list of floats\\
  flam\_err & Error on ``flam" & erg/s/\AA/cm$^2$ & list of floats\\
  sim\_flam & Input or true flux & erg/s/\AA/cm$^2$ & list of floats\\
\enddata
\end{deluxetable*}

\end{document}